\documentclass[reprint, amsmath,amssymb, prl,floatfix,superscriptaddress]{revtex4-2}

\usepackage{graphicx}
\usepackage{bm}
\usepackage{mathrsfs}  
\usepackage[usenames,dvipsnames]{xcolor}
\usepackage[colorlinks=true,citecolor=blue]{hyperref} 

\newcommand{\figref}[1]{Fig.\,\ref{#1}}

\begin{document}

\title{Real-space imaging of triplon excitations in engineered quantum magnets}

\author{Robert Drost}
\email{robert.drost@aalto.fi}
\affiliation{Aalto University, Department of Applied Physics, 00076 Aalto, Finland}

\author{Shawulienu Kezilebieke}
\affiliation{Department of Physics, Department of Chemistry and Nanoscience Center, 
University of Jyväskyl\"a, University of Jyv\"askyl\"a, 40014 Jyv\"askyl\"a, Finland}

\author{Jose L. Lado}%
\affiliation{Aalto University, Department of Applied Physics, 00076 Aalto, Finland}

\author{Peter Liljeroth}
\affiliation{Aalto University, Department of Applied Physics, 00076 Aalto, Finland}

\date{\today}

\begin{abstract}
Quantum magnets provide a powerful platform to explore complex quantum many-body phenomena. One example is triplon excitations, exotic many-body modes emerging from propagating singlet-triplet transitions. We engineer a minimal quantum magnet from organic molecules and demonstrate the emergence of dispersive triplon modes in one- and two-dimensional assemblies probed with scanning tunneling microscopy and spectroscopy. Our results provide the first demonstration of dispersive triplon excitations from a real-space measurement.

\end{abstract}
\maketitle

Quantum magnets provide a powerful platform to explore complex quantum many-body phenomena. In particular,
many-body excitations in quantum magnets provide a playground to engineer emergent
excitations only present in quantum materials \cite{sachdev2008, vasiliev2018, malki2020}. Well known examples are magnons \cite{chumak2015} in ordered magnets and spinons \cite{faddeev1981, kukushkin2009} in quantum spin liquid systems. Both spinons and magnons emerge due to fluctuations of localized magnetic moments, and are associated to emergent quasiparticles with $S=1/2$ or $S=1$ appearing in the material. These two excitations have remained at the core of quantum magnetism, with a variety of methodologies and experimental techniques focused on distinguishing these modes in real experiments\,\cite{hao2009, han2012, mourigal2013, dalla2015}. Quantum magnets can feature an additional more subtle type of excitation, emerging even when the individual building block of the quantum magnet is $S=0$: The triplons.

\begin{figure}[t!]
	\centering
		\includegraphics[width=1.00\columnwidth]{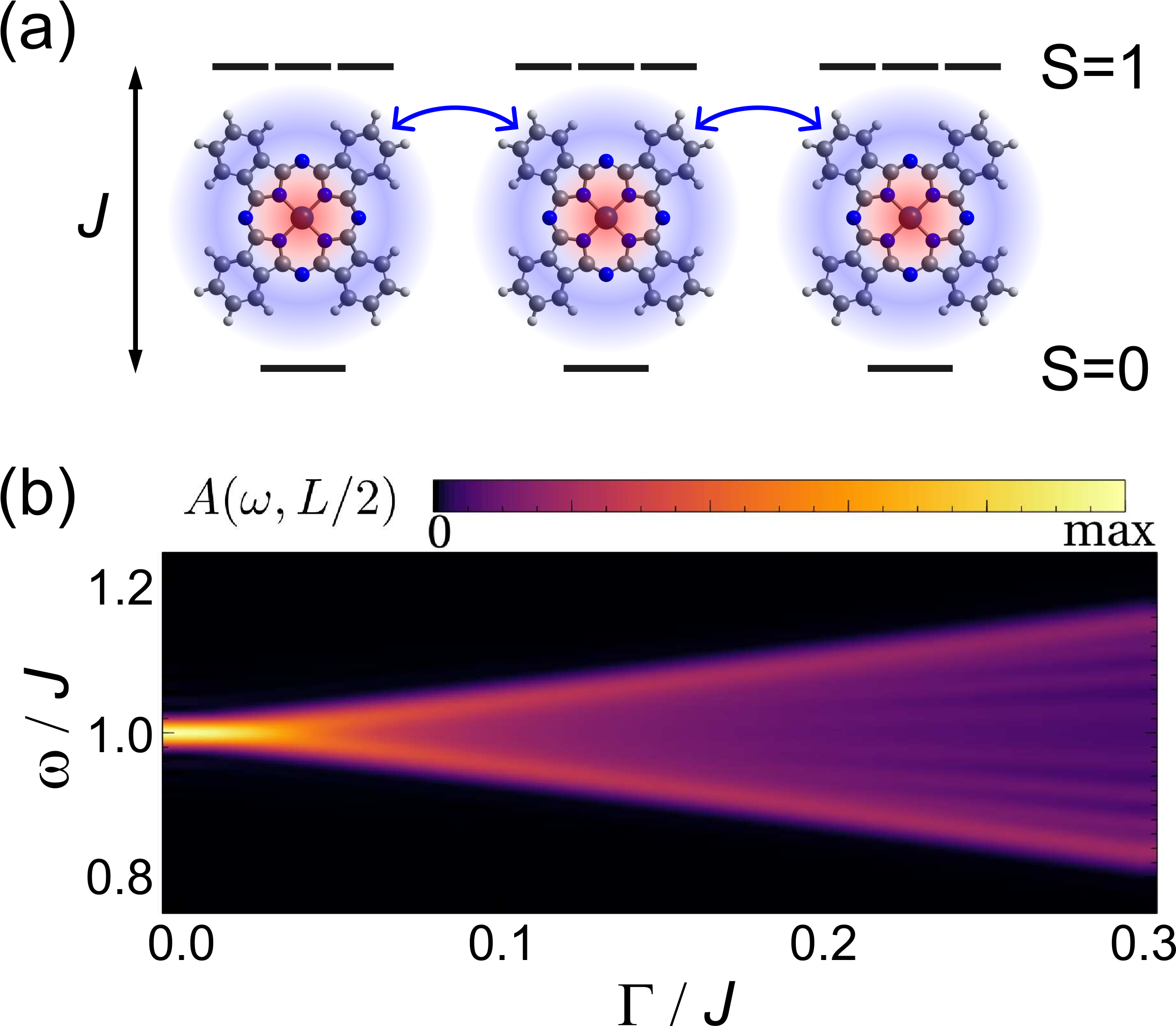}
	\caption{\textbf{Triplon excitations in a molecular spin chain.} (a) CoPC on NbSe$_2$ hosts two unpaired electrons in different molecular orbital. The molecule has a singlet ground state, in which the electronic spins are anti-parallel. Flipping one of the spins brings the molecule into the triplet state and creates a triplon. This fundamental excitation propagates through the chain through inter-molecular coupling. (b) Calculated triplon spectral function $A(\omega,L/2)$ for a molecular chain with $L=20$ sites in the middle of the chain, as a function of the inter-molecular coupling $\Gamma$.}
	\label{fig:Figure1}
\end{figure}

Triplon excitations emerge from an internal excitation of each building block of the quantum magnet, going beyond the picture applicable to spinons and magnons\,\cite{PhysRevB.41.9323,cavadini2001, xu2007}. When these internal singlet-triplet transitions in a quantum magnet become dispersive in the material, triplon modes emerge (see \figref{fig:Figure1}) \cite{McClarty2017,PhysRevLett.98.027403,Nawa2019}. Triplons are challenging to observe in conventional materials where either magnons or spinons are normally pursued. This stems from the fact that, for magnetic elements in bulk materials, the energy scales of singlet-triplet transitions is associated with Hund's energy, on the order of eV, and is dramatically larger than the typical bandwidth of spin fluctuations, on the order of meV \cite{coey2010magnetism}. Molecular systems \cite{Coronado2019}, however, provide natural magnetic systems with competing ground states, including singlet-triplet transitions. This intrinsic competition, in stark contrast to bulk compounds, makes engineered molecular systems natural platforms to explore triplon excitations. 

\begin{figure}[t!]
	\centering
		\includegraphics[width=1.00\columnwidth]{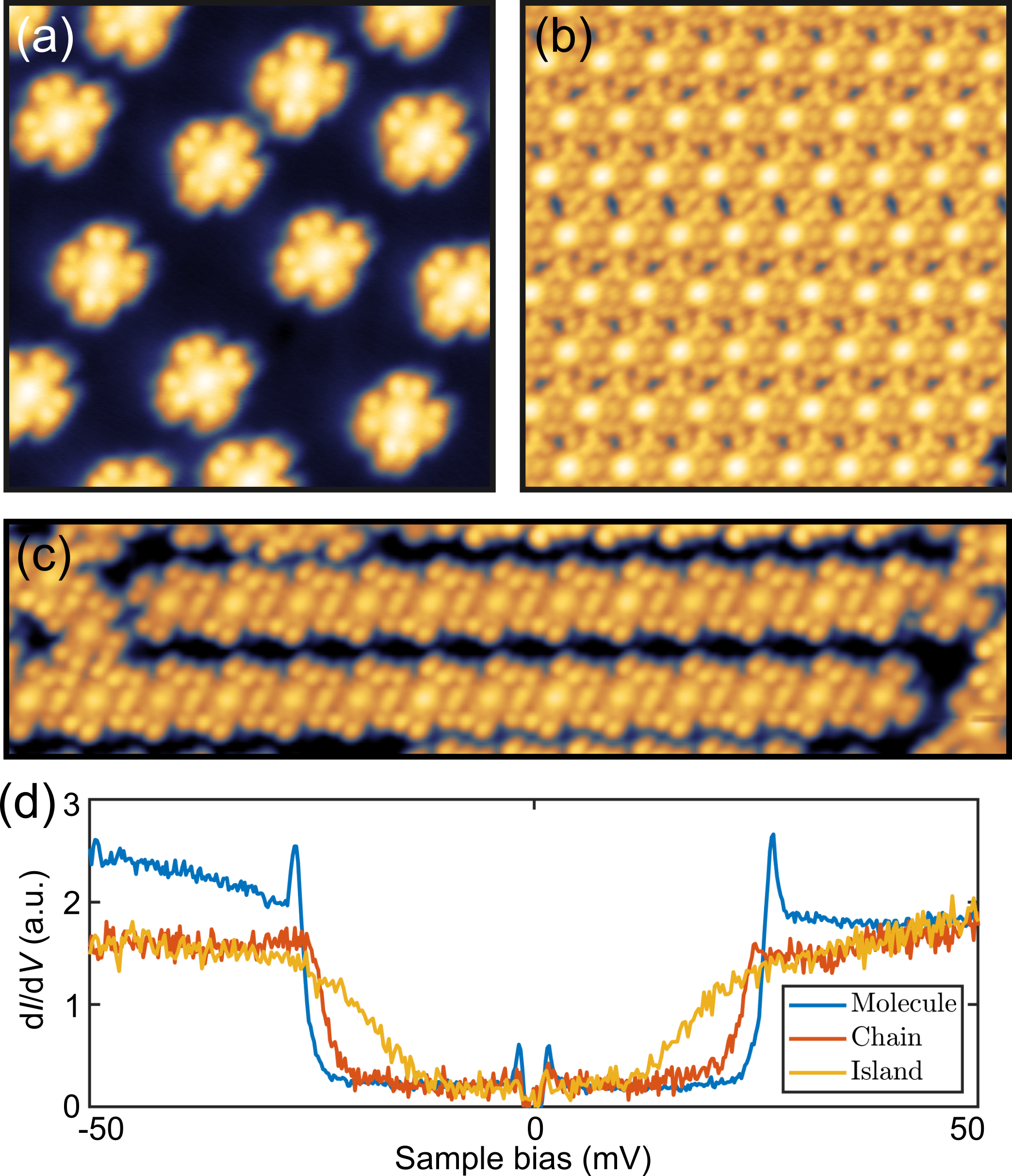}
	\caption{\textbf{CoPC molecules on NbSe$_2$.} (a) CoPC adsorbs as individual molecules at low coverage. Image size 10$\times$10\,nm$^2$. Increasing the surface coverage yields molecular islands (b) and chains (c) through self-assembly. Image sizes 12$\times$12\,nm$^2$ and 20$\times$5\,nm$^2$, respectively. (d) Conductance spectra acquired above the center atom of an individual CoPC (blue), a molecule in a chain (orange), and one in an island (yellow). The step-like features in the spectra correspond to spin excitations of the molecule by inelastic tunnelling.}
	\label{fig:Figure2}
\end{figure}

While triplon excitations have been studied in bulk compounds using e.g.~inelastic neutron scattering \cite{Kohno2007,PhysRevLett.100.205701,PhysRevLett.103.047401,RevModPhys.85.219,PhysRevLett.113.067201}, they have not been experimentally observed on the atomic scale. Here, we show that triplon excitations can be produced in designer quantum systems and be probed in real-space using scanning tunneling microscopy (STM). We achieve this using metal phthalocyanine molecules, where the spin ground state can be tailored through the nature of the central metal atom and choice of substrate \cite{Franke2011, malavolti2018, Kezilebieke2018,Kezilebieke2019,Wang2021}. A recent study found cobalt phthalocyanine (CoPC) on NbSe$_2$ to be in a singlet ground state, making it an ideal choice as a building block for structures with triplon excitations\,\cite{Wang2021}. By designing both one-dimensional and two-dimensional molecular arrays, we demonstrate that dispersive triplons can be measured with inelastic spectroscopy. We further show that the dispersion bandwidth of triplons is strongly correlated with the dimensionality of the molecular assembly as expected from dispersive many-body modes. Our results provide a real-space demonstration of triplon modes in engineered molecular systems, suggesting this as a potential platform for realizing exotic many-body phenomena experimentally.

\figref{fig:Figure2} shows low-temperature STM images of CoPC molecules on the NbSe$_2$ substrate. The sample preparation is described in detail in the Supplementary Information (SI). Briefly, we sublime CoPc molecules onto a freshly cleaved NbSe$_2$ substrate under ultra-high vacuum (UHV) conditions. Subsequently, the sample is inserted into a low-temperature STM operating at 4\,K housed within the same UHV setup. All the d$I$/d$V$ spectra shown here have been acquired with NbSe$_2$-coated superconducting tip \cite{Kezilebieke2018}. CoPC on NbSe$_2$ self-assembles into various motifs depending on the surface coverage. Low coverage samples yield individual molecules (\figref{fig:Figure2}a), while CoPC self-assembles into molecular chains (b) and islands (c) at higher coverage. As Wang and co-workers demonstrated recently\,\cite{Wang2021}, the molecule has two distinct adsorption sites on this surface. Depending on the alignment of the high-symmetry axes of molecule and substrate, it may display either Yu-Shiba-Rusinov states, or spin-flip excitations\,\cite{ternes2015,Kezilebieke2018,Kezilebieke2019,Wang2021}. 
Here, we focus exclusively on the latter kind, as our aim is to engineer novel magnetic phases from an $S=0$ ground state.
The spin moments arise from two unpaired electrons occupying the molecular orbitals of CoPC. The two molecular orbitals, one located at the center of the molecule and another located in in carbon atoms away from the center.
Molecules in one- and two-dimensional (1D, 2D) assemblies exclusively display spin-flip excitations.

\figref{fig:Figure2}d shows a representative conductance spectrum acquired above the metal center of an individual CoPC molecule in blue. Two features dominate the spectrum: The superconducting gap close to zero bias, and two sharp steps at higher energy. The superconducting gap arises from the NbSe$_2$ substrate and is not relevant for the spin-excitation features discussed here. The steps, placed symmetrically around the origin, are the characteristic signature of inelastic spin-flip excitations\,\cite{heinrich2004, hirjibehedin2006, Wang2021}. In this case, a tunnelling electron with sufficient energy excites the spin system into the triplet state. The spectra have been acquired with NbSe$_2$-coated superconducting tip \cite{Kezilebieke2018} to enhance the energy resolution beyond the thermal limit\,\cite{Pan1998,Franke2011}. This also results in a sharp peak at the edge of the spin-flip excitation steps arising from the superconducting density of states of the tip.\, \figref{fig:Figure2}d also shows spectra acquired on molecules in chains or islands (shown in orange and yellow, respectively). These show a similar step-like feature, which is, however, shifted to significantly smaller energies while being considerably broader than on single molecules. The effect becomes more pronounced as we move from 1D assemblies to 2D islands.

\begin{figure}[t!]
	\centering
		\includegraphics[width=1.00\columnwidth]{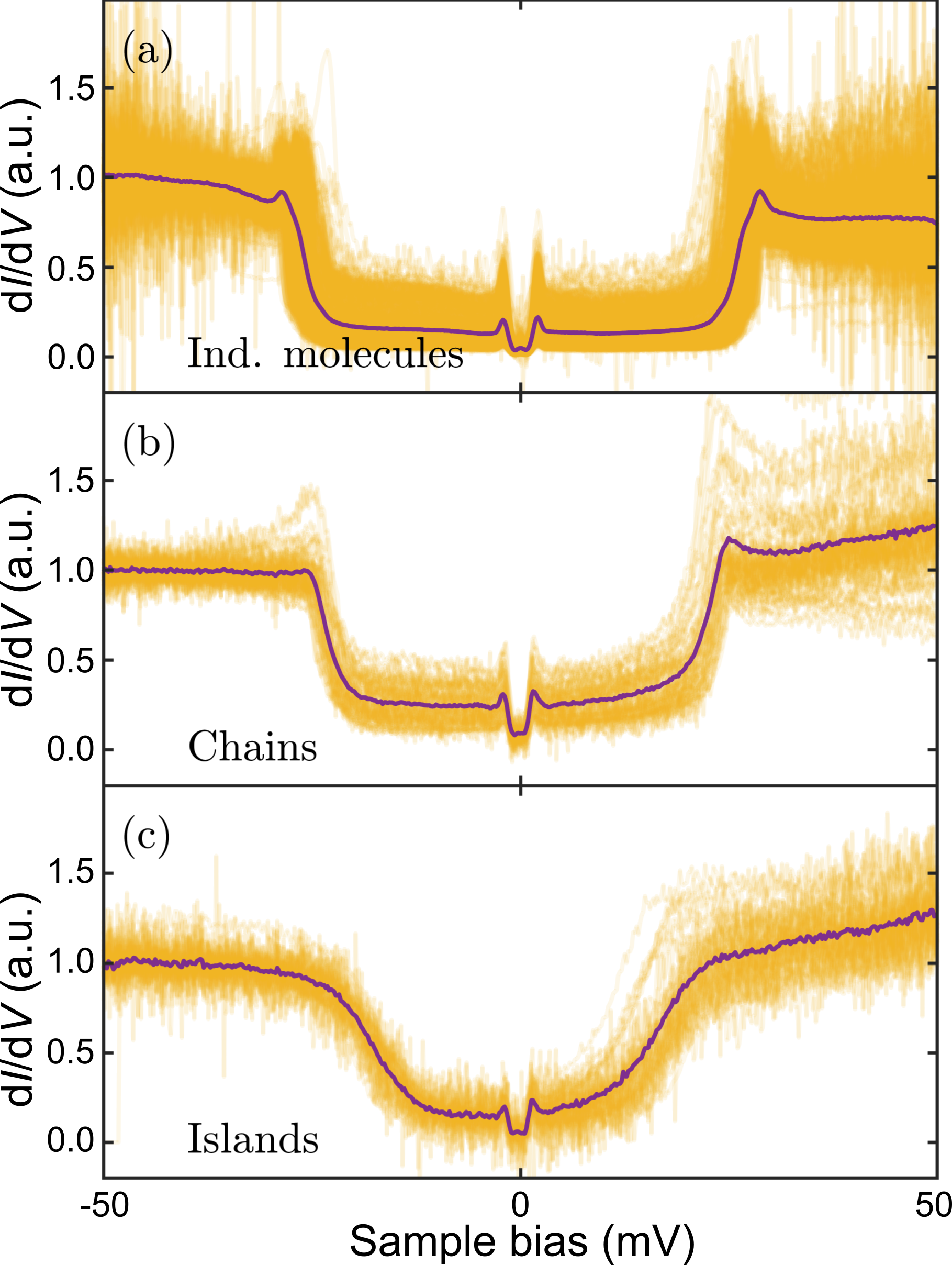}
	\caption{\textbf{Complete data set used in our analysis}. Conductance spectra acquired above (a) individual molecules, (b) molecules in chains, (c) molecules in islands, excluding edges. Purple lines are averages over all measurements in their respective class.}
	\label{fig:Figure3}
\end{figure}

The adsorption sites of CoPC are not commensurate with the NbSe$_2$ CDW reconstruction\,\cite{ugeda2016}. This creates locally different environments for each molecule, which has been previously shown to modulate the energies of YSR states of Fe atoms on NbSe$_2$\,\cite{Liebhaber2019}. We observe that the location of the inelastic excitation steps varies by a few mV from one molecule to the next. These changes are likely due to adsorption-site driven changes to the molecular orbitals, affecting the exchange coupling inside each molecule depending on its surroundings \cite{bork2011,kai2019}. This modulation in the step energy makes a quantitative analysis of the data on a molecule-by-molecule basis difficult. We therefore adopt a statistical approach and search for broader trends in a large data set, which is summarised in \figref{fig:Figure3}. The data set consists of 713 conductance spectra of CoPC on NbSe$_2$, which we group into three categories depending on the adsorption geometry: Individual molecules (\figref{fig:Figure3}a), 1D chains (\figref{fig:Figure3}b), and 2D islands (\figref{fig:Figure3}c).

\begin{figure}[t]
	\centering
		\includegraphics[width=1.00\columnwidth]{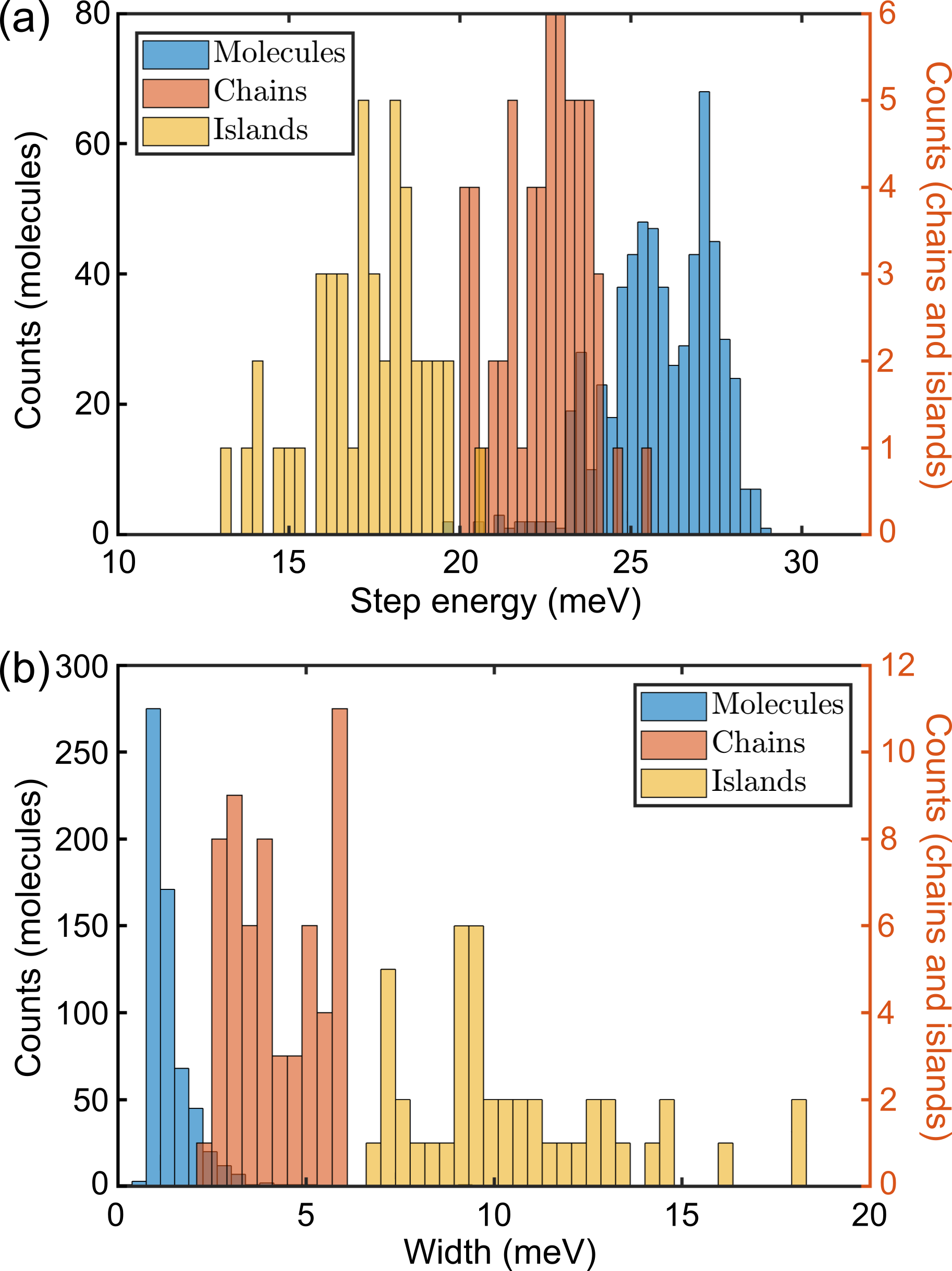}
	\caption{\textbf{Step energies and widths.} Inelastic step energy (a) and width (b) extracted by our fitting procedure for individual molecules (blue), molecules in chains (orange), and islands (yellow). The spin excitation step moves to lower energy and broadens as the number of neighbour molecules increases.}
	\label{fig:Figure4}
\end{figure}

We quantify the changes in the spectra by characterizing each spectrum by the location of the step and its width, by fitting them with a Fermi step function mirrored about the origin\,\cite{SI}\nocite{PhysRevLett.69.2863, PhysRevResearch.1.033009, PhysRevResearch.2.023347, PhysRevB.90.045144, PhysRevB.90.115124, RevModPhys.78.275, 2022arXiv221207893K}. This method reliably extracts the centre of the inelastic step, while the effective temperature of the fit function encodes all broadening mechanisms present in the experiment. The results of our analysis are summarised in \figref{fig:Figure4}. They quantify the trends already evident from the conductance spectra shown in \figref{fig:Figure3}. Individual molecules display by far the largest spin excitation gap of all molecular species. The multi-peak distribution is likely due to interactions with the CDW. The excitation gap is significantly reduced for molecules in one- and two-dimensional assemblies (see \figref{fig:Figure4}a). The more close neighbours surround a molecule, the lower the spin excitation energy becomes. 

The position of the inelastic step may be influenced by a multitude of factors, and in particular solely by changes in the local environment \cite{wang1993,gambardella2009,heinrich2015}.  A shift of the step energy is therefore not related with inter-molecular spin-spin interactions. In addition to the step energy, the width of the inelastic step, shown in \figref{fig:Figure4}b increases when going from molecules to chains and to islands. While individual molecules generally show values close to the expected thermal broadening at $T= 4.2$\,K, data from one- and two-dimensional assemblies yields significantly larger values of the step width. The effect is more pronounced the more neighbours surround a given molecule.

Changes in the local adsorption geometry alone cannot explain this broadening of the inelastic step. Factors like anisotropy or other environmentally driven changes to the intra-molecular exchange interaction do not affect the width of an excitation feature. The clear and consistent broadening of the inelastic step observed in our experiment is direct evidence of new inter-molecular interactions and the formation of spin chains and lattices. The existence of these new couplings implies that excitations produced at any point in a structure can traverse it as triplons.

To support our findings, we build a model for spin chains mimicking the situation in our experiment. CoPC on NbSe$_2$ hosts two magnetic moments in different molecular orbitals \cite{Wang2021}. One of those, labeled $\bm{K}$, is located on the centre ion, the second, labeled $\bm{S}$, distributed over the outer ligands. Each site of our spin chain hence consists of a pair of spin operators $\bm{K}$ and $\bm{S}$ interacting through an intra-molecular exchange coupling $J$. The inter-molecular exchange coupling $\Gamma$ is driven by the spatial overlap of molecular orbitals at adjacent sites. The ligand orbitals will be the primary sources of inter-molecular coupling (see sketch in  \figref{fig:Figure1}a), leading to a Hamiltonian of the form:

\begin{equation}
    \mathscr{H} =  \sum_i J^{}_i \bm{S}^{}_i \cdot \bm{K}^{}_{i} + 
    \Gamma \sum_{\langle i, j \rangle} \bm{S}^{}_i \cdot  \bm{S}^{}_j
    \label{eq:ham}
\end{equation}

Here, $\bm{S}^{}_i$ and $\bm{K}^{}_i$ are spin-1/2 operators at lattice site $i$, $J^{}_i$ is the intra-molecular exchange coupling at site $i$, $\Gamma$ the inter-molecular exchange coupling, and the sum $\left<i, j\right>$ is taken over nearest neighbours. The spin-spectral function of the model $A(\omega,n) = \langle GS | K^z_n \delta(\omega - \mathscr{H}+E_{GS}) K^z_n | GS \rangle$ accounts for
the singlet-triplet transition signaled by the inelastic step, and in particular is related with the differential conductance as d$I/$d$V(V,n) \sim  \int_0^V A(V',n) dV'$. We solve the previous model using a tensor-network method and compute the spectral function with the kernel polynomial methodology \cite{dmrgpy,ITensor,10.21468/SciPostPhysCodeb.4}. Further details of the modelling are given in the the SI.

\begin{figure}[t!]
	\centering
		\includegraphics[width=1.00\columnwidth]{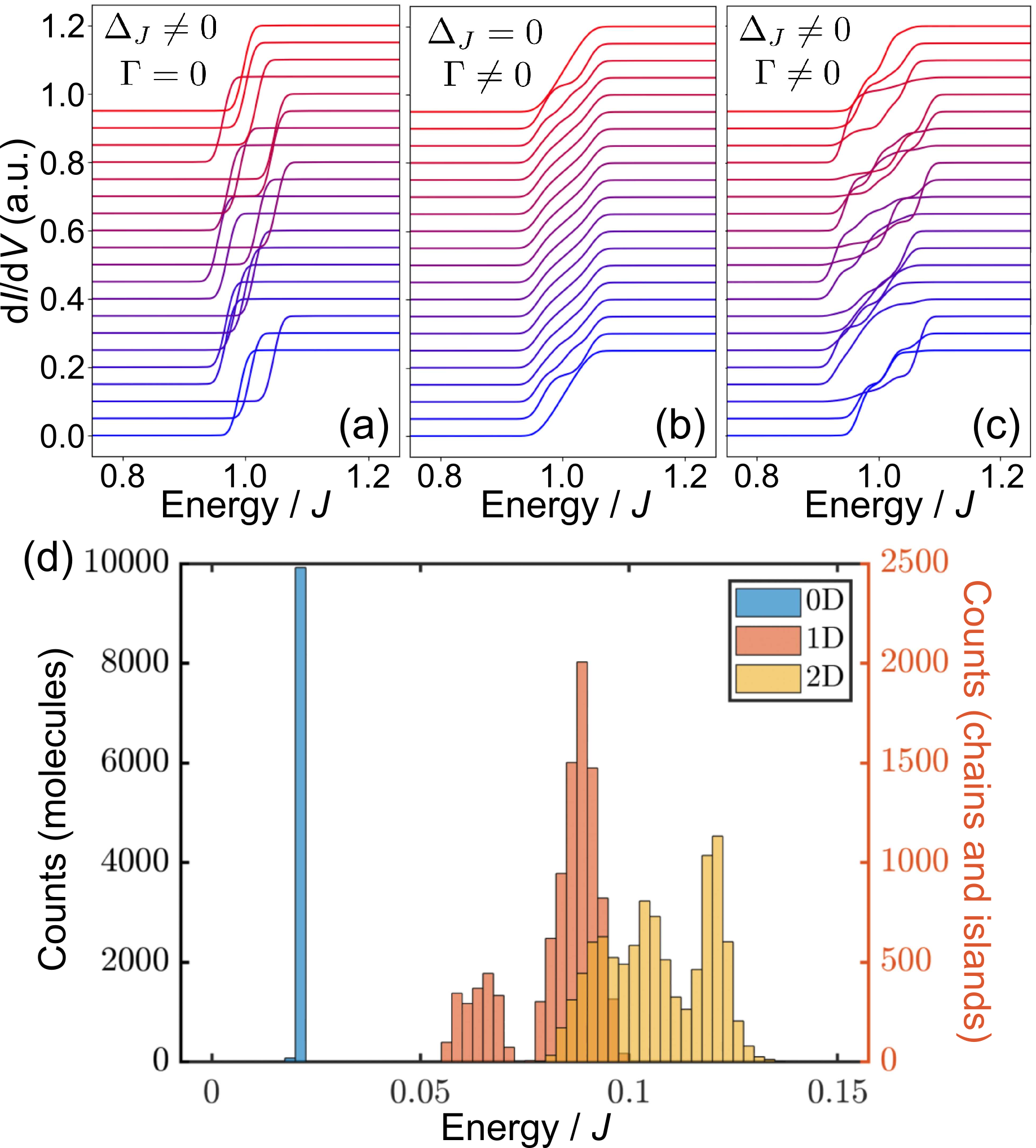}
	\caption{\textbf{Results from a spin model.} (a-c) Simulated d$I$/d$V$ curves
 in the absence of inter-molecule coupling (a), and for coupled one-dimensional chain in the absence of disorder (b), and with disorder (c).
 (d) Distribution of the step widths obtained from our model, given in units of the exchange coupling $J$. The multi modal distributions in the 1D and 2D cases arise from the presence of edge and corner atoms.
 We took $\Gamma = 0.1 J $ in (b,c) (see Eq.~(1)) and $\Delta_J = 0.1 J$ in (a,c).}
	\label{fig:Figure5}
\end{figure}

We simulate our experiment by computing the spin excitation spectrum at each lattice site and extracting the width of the resulting step. The distributions of step energies and widths is directly comparable to our experimental results, and we model the CDW-induced disorder by introducing random variations of the intra-molecular coupling.
In its simplest form, with $J^{}_i = J$ and $\Gamma = 0$, the model produces a singlet-triplet excitation of energy $J$ at each lattice site. This situation corresponds to an idealized scenario of individual molecules with no disorder. Keeping $\Gamma = 0$, but allowing for randomized variations of $J^{}_i \in [J-\Delta_J/2,J+\Delta_J/2]$, produces shifts of the excitation energy between different sites as shown in \figref{fig:Figure5}a. This stems from the fact that the singlet-triplet splitting is uniquely determined by the exchange Hund's coupling. This configuration (\figref{fig:Figure5}a) represents measurements on individual molecules in our experiment, where the width of the step remains unaffected by the variations of $J^{}_i$. These results indeed reproduce with good accuracy the situation in our experiment: A broad distribution of excitation energies, but relatively uniform step widths (see \figref{fig:Figure4}a).

We now move on to the case with a finite coupling $\Gamma \ne 0$, considering first a pristine molecular chain with uniform Hund's coupling $J$ (\figref{fig:Figure5}b).  In the bulk of the chain, as shown in \figref{fig:Figure5}b, the broadening of the step converges to a fixed value dominated by the inter-molecular coupling $\Gamma$. The coupling $\Gamma$ acts as an effective broadening mechanism and increases the width of the step. This increased broadening accounts for the emergence of a triplon band\cite{PhysRevB.41.9323}. The experimentally observed limit in the chains corresponds to a finite coupling $\Gamma$ in the presence of small disorder in $J$, shown in \figref{fig:Figure5}c. This limit is directly accounted for by taking $\langle J_i \rangle = J$ and $J_i \in [J-\Delta_J/2,J+\Delta_J/2]$.  It is observed that the broadening of the steps remains, as in the pristine limit, but now with a slight shift of its onset driven by the different $J_i$.

We now evaluate our results using a statistical approach, considering both isolated molecules, chains and island as in the experiment. \figref{fig:Figure5}d shows the step width obtained in our model in different lattice geometries. The simulated distributions are in good agreement with the experimental results: Starting from a narrow distribution for individual molecules, both the mean and the spread of values increase in the case of chains, and even further for islands. The simulations thus reproduce all essential aspects of our experiment. The increasing step width is directly related to a finite inter-molecular spin-spin coupling, which allows to extract a typical value of $\Gamma \approx 8$ meV. The successive broadening of the inelastic feature we observe in tunnelling spectroscopy is thus direct evidence of not only triplon formation, but also their propagation within molecular chains and islands. Arrays of CoPC on NbSe$_2$ thus form a minimal working example of a quantum magnet with a characteristic gapped spin excitation above the ground state. The model described by Eq.~(1) is known to describe the effective spin dynamics of the Kondo insulators \cite{PhysRevB.96.075115,PhysRevB.99.235130,2023arXiv230209701C}, tuning the intermolecular interaction appropriately would enable the study of the quantum phase transition from the Kondo singlet to ordered antiferromagnetic phase in these systems.

The hunt for quantum magnets and exotic magnetic excitations has been mostly focused on magnetic insulators with complicated material chemistry and crystal structures. Our experiments show that arrays of metal-organic molecules are efficient platforms to simulate quantum magnets and study their excitations in a simplified setting. Magnetic molecules are adaptable model systems in which the magnetic interactions and ground state can be tuned through organic chemistry. Our work shows that they can form the building blocks of designer quantum magnets, offering a pathway to creating, exploring, and ultimately understanding magnetic excitations in quantum matter.

\begin{acknowledgements}
   
This research made use of the Aalto Nanomicroscopy Center (Aalto NMC) facilities and was supported by the European Research Council (ERC-2021-StG no.~101039500 ``Tailoring Quantum Matter on the Flatland'' and ERC-2017-AdG no.~788185 ``Artificial Designer Materials'') and Academy of Finland (Academy professor funding nos.~318995 and 320555, Academy research fellow nos.~331342, 336243, 338478, 346654, 347266, and 353839). We acknowledge the computational resources provided by the Aalto Science-IT project. We thank Teemu Ojanen, Yuqi Wang and Markus Ternes for fruitful discussions during this project.
\end{acknowledgements}

%

\newpage
\phantom{some sample text}
\newpage

\setcounter{figure}{0}
\setcounter{equation}{0}
\newcommand{\suppcite}[1]{\cite[S\hspace{-1.8mm}][]{#1}}
\renewcommand\thefigure{S\arabic{figure}}
\renewcommand\theequation{S\arabic{equation}}

\makeatletter
	\renewcommand*{\@biblabel}[1]{[S#1]}
\makeatother

\begin{center}
\onecolumngrid
\section{Supplementary Information for 'Real-space imaging of dispersive triplon excitations in engineered quantum magnets'}
\vspace{2cm}
\end{center}

\twocolumngrid

\section{Experimental details}

All data shown was acquired in a Unisoku USM1300 STM operating at 4\,K. High-quality NbSe$_2$ crystals were obtained from HQ Graphene. We glue our substrates directly onto a metallic support plate with conductive epoxy (Epotek H21D) to achieve a reliable electrical contact. We prepare samples by applying commercial scotch tape to NbSe$_2$ in the ambient and cleaving the crystals by peeling off the tape in the load-lock chamber of our ultra-high vacuum (UHV) system. We ascertain the quality and cleanliness of the cleaved crystal faces by STM measurements. CoPC is sublimed in UHV from a crucible kept at 660 K onto the sample at kept at room temperature. We control the surface coverage by adjusting the deposition time. CoPC on NbSe$_2$ self-assembles into various motifs depending on their total surface coverage. Low coverage samples yield individual molecules, while CoPC self-assembles into molecular chains and islands at higher coverage. All the d$I$/d$V$ spectra shown here have been acquired with NbSe$_2$-coated superconducting tip \suppcite{Kezilebieke2018}. 

\section{Determination of IETS step energy and width}

\begin{figure}[!h]
	\centering
		\includegraphics[width=1.00\columnwidth]{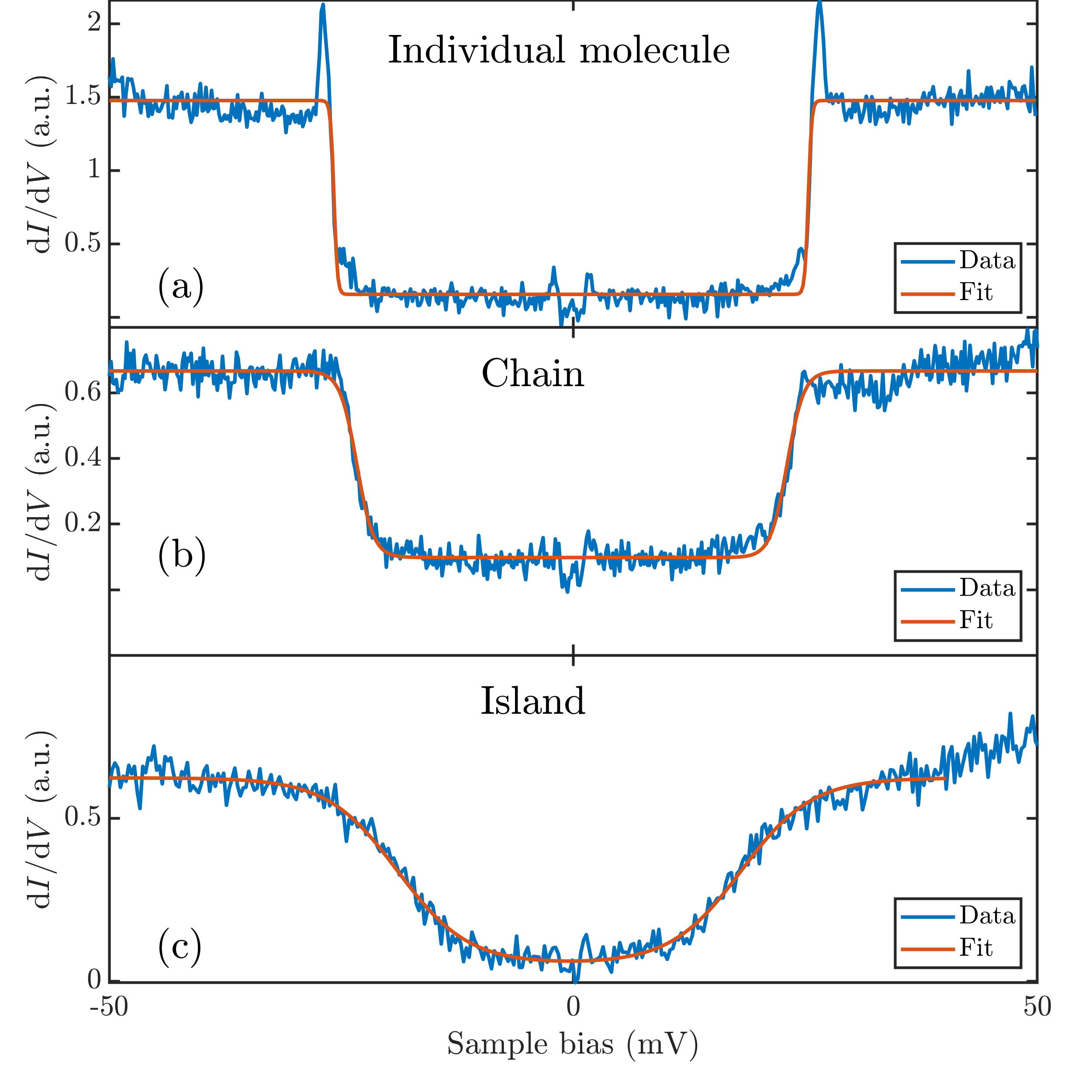}
	\caption{Representative data from (a) and individual CoPC molecule, (b) a CoPC molecule in a self-assembled chain, and (c) a molecule in an island with fits using the model in Eq. \ref{eq:eq1}.}
	\label{fig:FigureS1}
\end{figure}

Our approach is intended to be agnostic to locally changing parameters, such as the magneto-crystalline anisotropy, and instead focuses on statistical trends in the step width. We therefore analyse our data by fitting conductance spectra using a phenomenological model. The main feature in our data are the two step-like increases of the tunnelling conductance. If arising from spin excitations, these must be symmetrical around zero bias. We do not consider supercocnductivity in the tip or sample in our analysis. The energy of the superconducting gap in NbSe$_2$ ($\Delta\approx 1.4$\,mV) is much smaller than either the spin excitation energy or the changes thereof and only contributes a rigid shift. We  model the data using a sum of two Fermi functions

\begin{eqnarray}
    G(\epsilon) & = & A \left( \frac{1}{\exp \left[-(\epsilon_s + \epsilon)/k_B T^{}_{\text{eff}}\right] + 1} \right. \nonumber \\
        & & \left. \hspace{3mm} + \frac{1}{\exp \left[-(\epsilon_s - \epsilon)/k_B T^{}_{\text{eff}}\right] + 1} \right) + B,
        \label{eq:eq1}
\end{eqnarray}
where $A$ is the step height, $\epsilon_s$ is the step energy, $k_B$ the Boltzman constant, $T^{}_{\text{eff}}$ and effective temperature combining all broadening mechanisms in the experiment, and $B$ the the spectral baseline. As shown in \figref{fig:FigureS1}, this model fits the data in all cases concerned in the experiment and is successful in extracting the key parameters for our analysis.

We use the concept of an effective temperature to evaluate the width of the IETS steps observed in the experiment. The inter-molecular spin-spin coupling leads to a splitting of the single IETS step into several closely spaced spin excitations. As these cannot be individually resolved at our experimental temperature, they blend into each other, thus producing a broad spectra feature. We think of this as an additional, and indeed the dominant, broadening mechanism, which can be absorbed into $T^{}_{\text{eff}}$. The quality of the fits obtained with the phenomenological model in Eq.~\ref{eq:eq1} suggests that this is a valid approach in the context of our data.

\begin{figure}[!h]
	\centering
		\includegraphics[width=1.00\columnwidth]{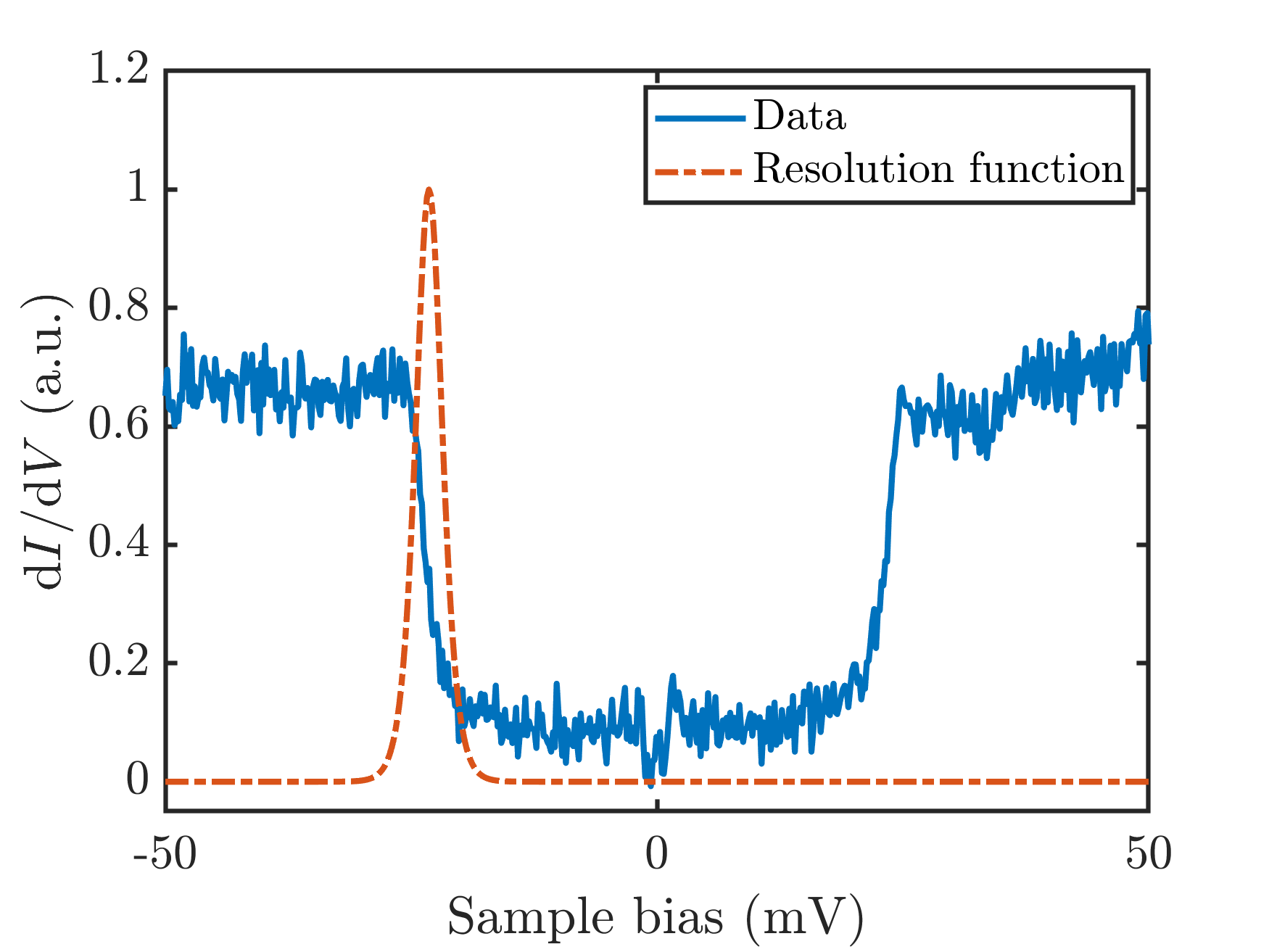}
	\caption{Experimental data from a CoPC molecule in a self-assembled chain in blue. Superimposed at the IETS step energy is the derivative of the Fermi function at $T^{}_{\text{eff}}$, which we use to evaluate the width of the IETS feature.}
	\label{fig:FigureS2}
\end{figure}

When modelling data, the effect of temperature related broadening is most easily accounted for by calculating conductance spectra at zero temperature and convolution them with the derivative of the Fermi function evaluated at the electronic temperature of the experiment:
\begin{equation}
    G(T) = G(T=0)*\frac{1}{k_B T} \frac{\exp{(-\epsilon/k_B T)}}{\exp{(-\epsilon/k_B T) + 1}}
\end{equation}
The derivative of the Fermi function can thus be regarded as a resolution function of the experiment. We follow a similar approach here and take the full width at half maximum of the derivative of the Fermi function at the effective temperature as the width of the IETS step. While the premise of a single sharp feature in the zero temperature limit does not apply to our data, this approach still provides an objective and reliable estimation of the width. For an illustration, see \figref{fig:FigureS2}.

\section{Adsorption structure of CoPC on NbSe$_2$}

\begin{figure}[!h]
	\centering
		\includegraphics[width=1.00\columnwidth]{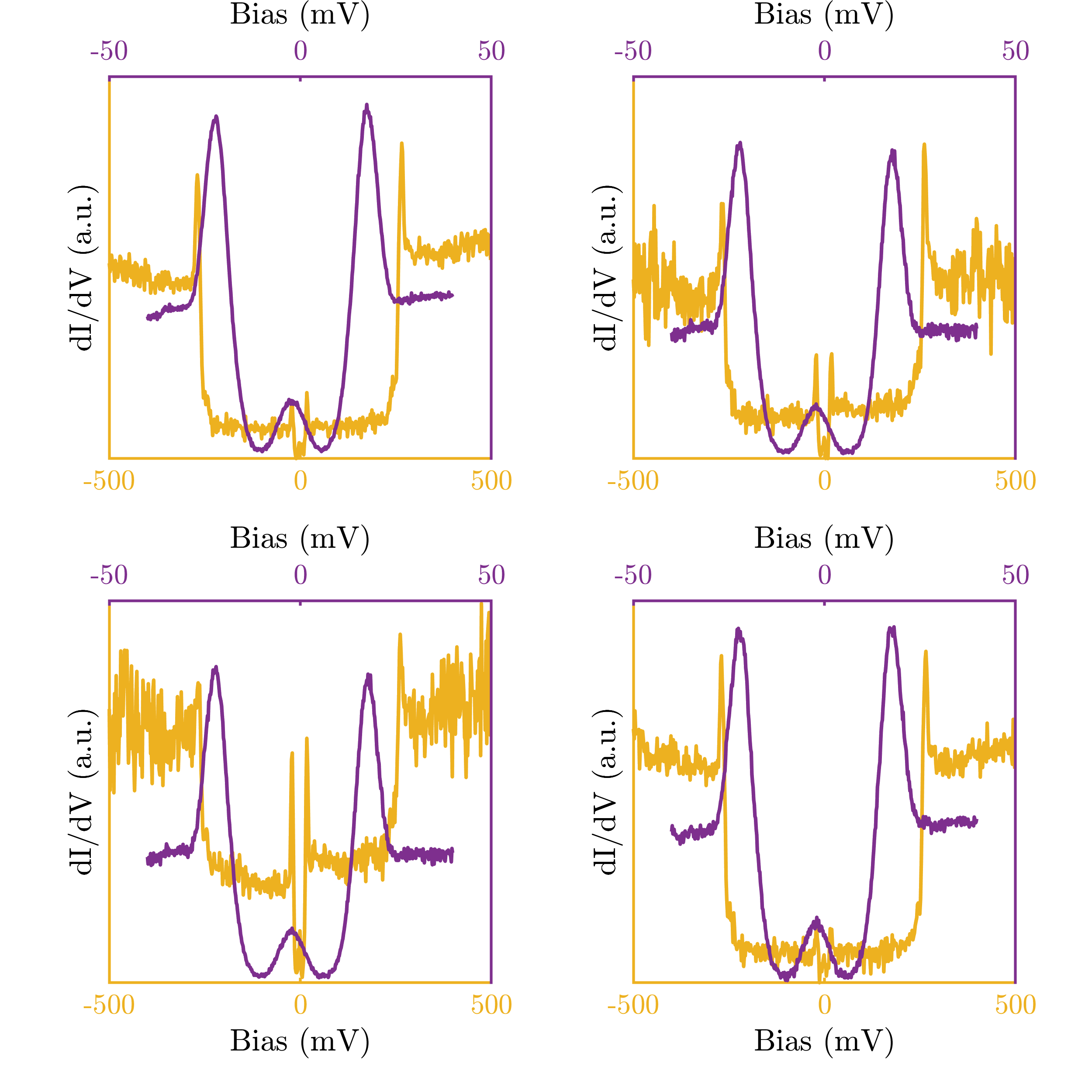}
	\caption{Conductance spectra of the superconducting gap (purple) and IETS steps (yellow) acquired above four different CoPC molecules. The absence of sub-gap states suggests the molecule in in the singlet ground state.}
	\label{fig:FigureS3}
\end{figure}

Based on the presence of IETS steps, as reported by Wang and co-authors \suppcite{wang2021}, we conclude that CoPC molecules adsorb with their molecular high-symmetry axis parallel to that of the substrate. This configuration leads to strong coupling and charge transfer with the substrate, resulting in a total molecular spin of $S = 0$, which is confirmed by measurements in varying magnetic field \suppcite{wang2021}. Our experiment reproduces the zero-field results published by Wang and co-workers. In particular, CoPC molecules which exhihbit IETS steps do not show any sub-gap states inside the superconducting gap (see \figref{fig:FigureS3}) that are associated with spinful impurities. We conclude that individual CoPC molecules are indeed in the $S=0$ ground state.

Molecules in 1D and 2D structures show no rotation with respect to each other. Inter-molecular distances in 1D structures are very close to four NbSe$_2$ lattice constants. 2D structures do not form close packed arrays and the inter-molecular distances are close to four or five times the NbSe$_2$ lattice constants. As all molecules further display IETS steps, we conclude that CoPC remains commensurate with the NbSe$_2$ crystal lattice, but \textit{not} the 3$\times$3 CDW reconstruction, and retains the $S=0$ ground state.

\begin{figure}[t!]
	\centering
		\includegraphics[width=\columnwidth]{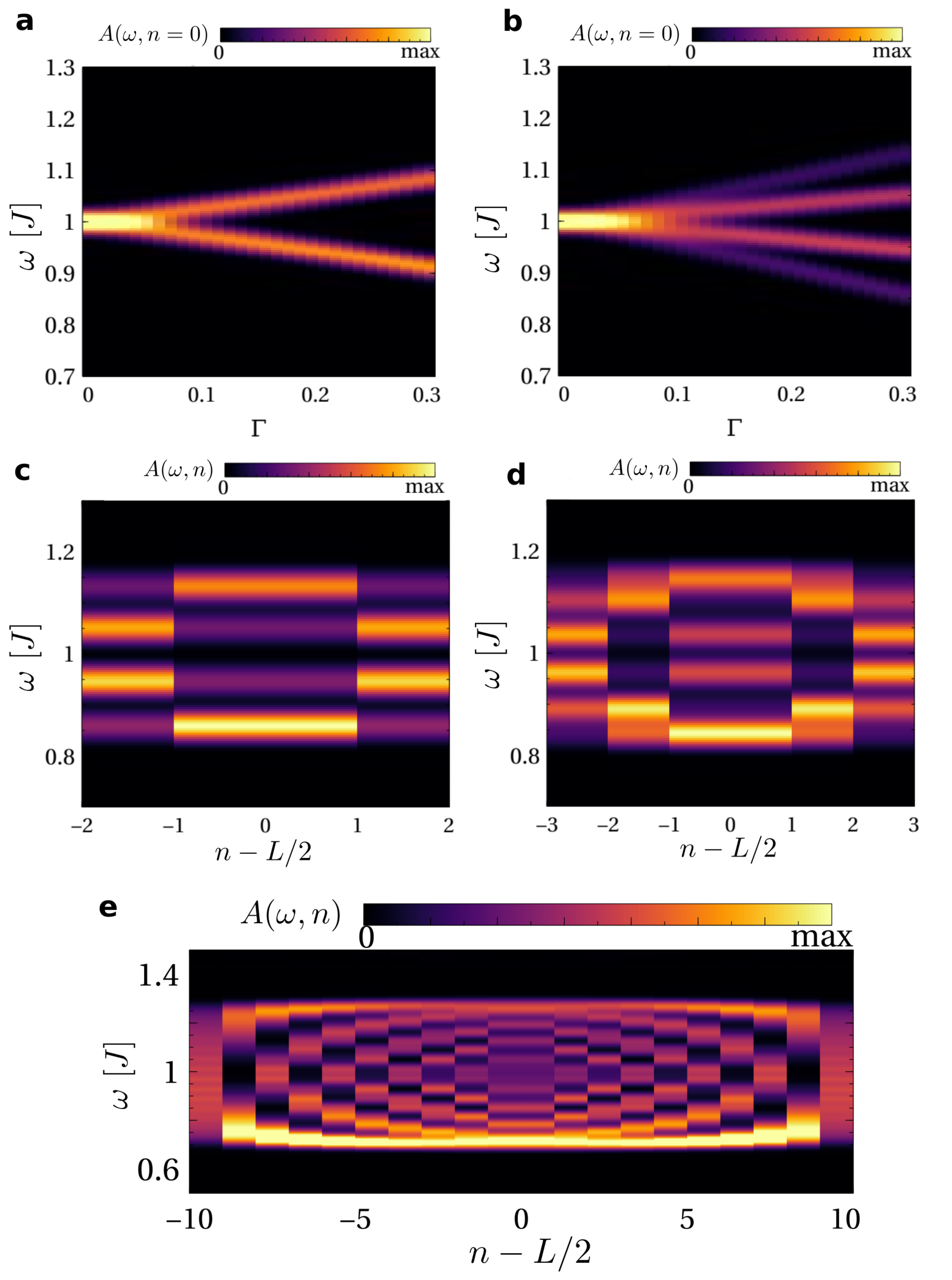}
	\caption{(a) Spin structure factor showing the singlet-triplet transition in a molecule as a function of the inter-molecule coupling $\Gamma$ for a dimer (a) and an $L=4$ chain (b). Panels
 (c,d) show the spatially resolved spin structure factor for a chain with $L=4$ (c) and $L=6$ (d), 
 highlighting the emergence of triplon excitations.
 We took $\Gamma=0.3J$ for (c,d) and $\Gamma=0.5J$ for (e).
 }
	\label{fig:smfig3}
\end{figure}

\section{Spin chain model}

Here we elaborate on the emergence of triplon excitations in the spin chain model.
As stated in the main manuscript, each molecule is formed by two $S=1/2$ coupled
antiferromagnetically with strength $J_i$, with the different molecules having a smaller
coupling $\Gamma$

\begin{equation}
    \mathscr{H} =  \sum_i J^{}_i \bm{S}^{}_i \cdot \bm{K}^{}_{i} + 
    \Gamma \sum_{\langle i, j \rangle} \bm{S}^{}_i \cdot  \bm{S}^{}_j
\end{equation}

For the sake of simplicity, we will first address the case $J^{}_i \equiv J$ first, namely when all the
Hund's couplings in the molecules are the same. In the absence of intermolecule coupling ($\Gamma = 0$),
each site hosts a singlet-triplet excitation at an energy $J$. The singlet-triplet excitation directly emerges when computing the dynamical spin structure factor

\begin{equation}
A(\omega,n) = 
\langle GS | K^z_n \delta(\mathscr{H} - \omega +E_{GS}) K^z_n | GS \rangle
\end{equation}
In an experimental measurement, the dynamical spin structure factor is proportional to the
$d^2I/dV^2$. For small systems 
$N \leq 10$), 
the previous Hamiltonian and spectral function
can be solved with exact diagonalization, whereas for large systems we use a tensor-network formalism \suppcite{dmrgpy,ITensor,10.21468/SciPostPhysCodeb.4}. For large systems, 
the ground state is solved using the density-matrix renormalization group
method with matrix-product states \suppcite{PhysRevLett.69.2863,ITensor,10.21468/SciPostPhysCodeb.4}, 
and the dynamical correlator is computed using a kernel polynomial tensor-network
method \suppcite{dmrgpy,PhysRevResearch.1.033009,PhysRevResearch.2.023347,PhysRevB.90.045144,PhysRevB.90.115124,RevModPhys.78.275}. The smearing of the tensor-network calculations is controlled with the number of Chebyshev polynomials in the expansion \suppcite{RevModPhys.78.275}.

\begin{figure}[t!]
	\centering
		\includegraphics[width=\columnwidth]{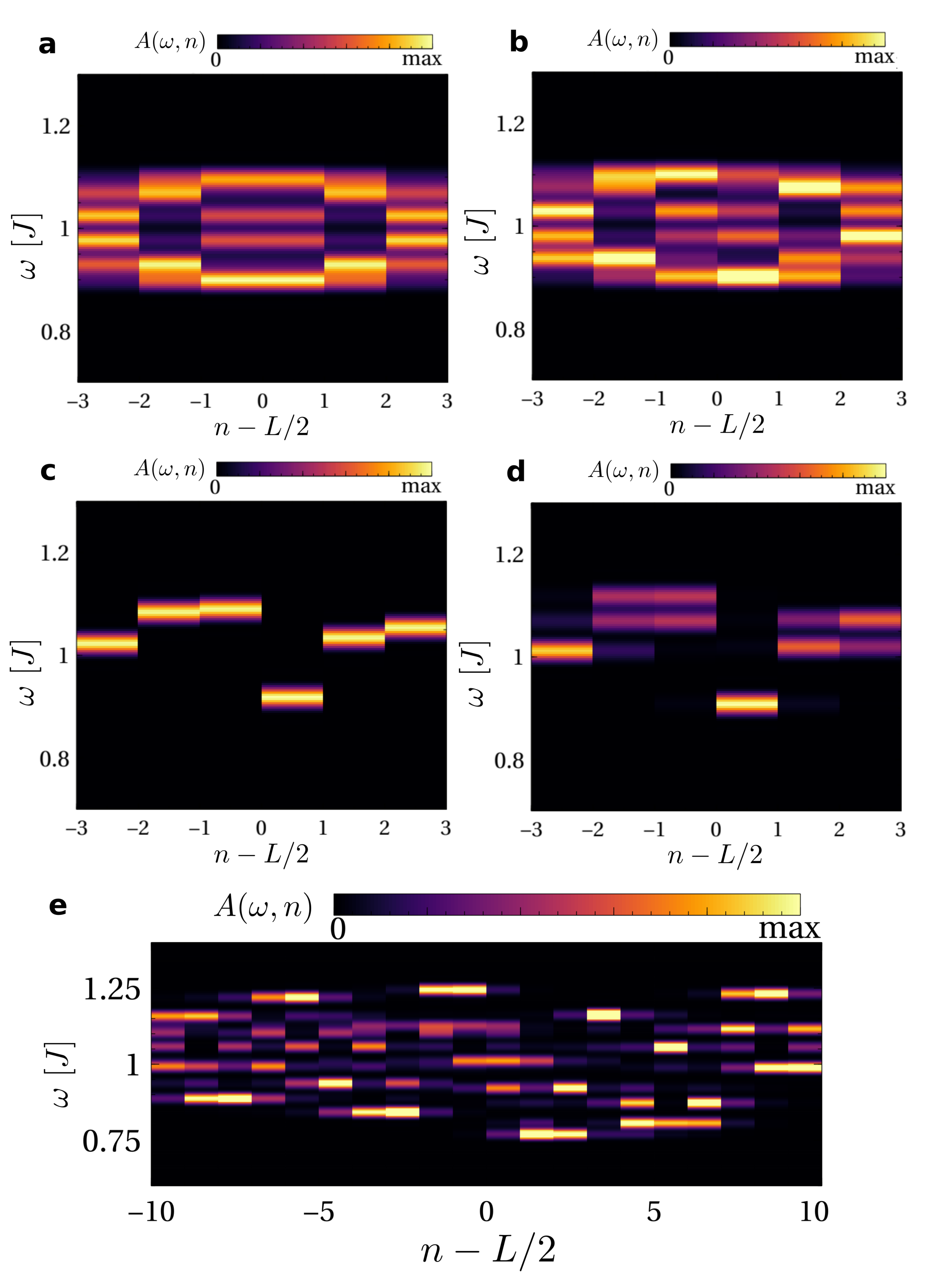}
	\caption{Spin structure factor in a coupled molecular chain
 without disorder (a), with weak disorder (b).
 Panels (c) shows the limit of Hund's disordered molecules that are disordered and decoupled (c), and disordered with weak coupling (d). 
 Panel (e) shows the case in which the disorder and coupling are on the
 same order.
 We took $\Delta_J=0.2J$ for (c,d), $\Delta_J = 0.03J$ for (b), $\Gamma = 0.2J$ for (a,b),
 $\Gamma = 0.1J$ for (d), $\Gamma = \Delta_J = 0.3J$ for (e).
 }
	\label{fig:smfig4}
\end{figure}

\begin{figure}[t!]
	\centering
		\includegraphics[width=\columnwidth]{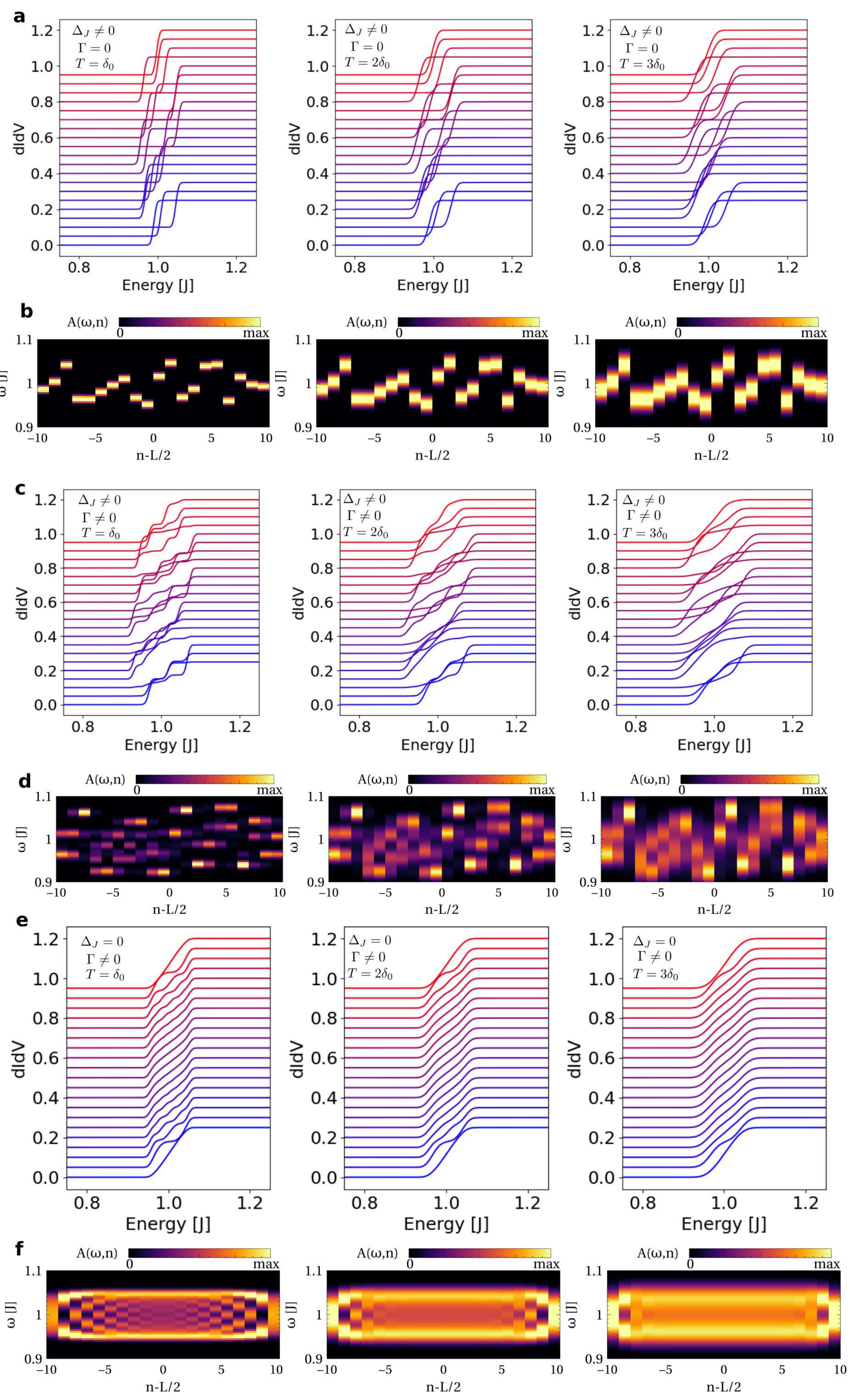}
	\caption{Disconnected disorder molecular chain (a,b), disordered coupled chain (c,d)
 and pristine coupled chain (e,f), showing both
 the simulated dIdV (a,c,e) and spin-structure factor (b,d,f) for different smearing. It is observed
 that the sequence of steps are washed away by the smearing as observed experimentally, yet maintaining a qualitative difference between isolated molecules and chains. The same disorder profile was takes for all the calculations. We took $\Delta_J = 0.1 J$ for (a,b,c,d) and $\Gamma = 0.1 J$ for (c,d,e,f).
 }
	\label{fig:smfig5}
\end{figure}

We show in \figref{fig:smfig3}a,b the local spin structure factor as a function of the coupling between molecules. As it is observed, the coupling $\Gamma$ creates a splitting in the triplon modes. As the
chain become larger (\figref{fig:smfig3}c-e), a set of dispersive triplon modes appear
in the chain, leading to a band with width $\Gamma$.

In our experiment, the underlying NbSe$_2$ substrate leads to a slightly different Hund's coupling in each molecule, that generates a different $J$ in each molecule. This phenomenology can be accounted by making the
$J$ spatially dependent $J_i$. For the sake of generality, we will take that $J_i$ takes a value
$J_i = J + \chi_i \Delta_J$, where $J$ is its average value, $\Delta_J$ is the typical modulation,
and $\chi_i$ is a random variable between $(-0.5,0.5)$. In the presence of such a Hund's disorder, the
onset for the triplon excitation will change from site to site. However, even in this disordered case, the
emergence of triplon dispersion can be observed. To highlight this, we show in \figref{fig:smfig4} the evolution of the triplon dispersion with and without coupling and disorder for the molecular chain.
It is observed that the dispersive states observe in the pristine chain (\figref{fig:smfig4}a) remain
in the presence of a weak Hund's disorder (\figref{fig:smfig4}b). In the absence of coupling between the molecules (\figref{fig:smfig4}c), individual excitation at different energies are present, and the inclusion of a small coupling leads to dispersion (\figref{fig:smfig4}d). As the coupling between molecules becomes stronger (\figref{fig:smfig4}e), the bandwidth of the triplet modes in each site becomes
larger even in the presence of disorder. Finally, it is worth noting that the substrate can induce a slight different $\Gamma$ coupling between different molecules. This modulation in $\Gamma$ would induce subtle changes in the triplon bandwidth on each site, yet with a sub-dominant effect in comparison with the modulation in $J$.

The impact of the broadening in the calculations is shown in \figref{fig:smfig5}. In particular, we focus
in the case with disorder, targeting both the limit of decoupled molecules $\Gamma=0$ and with 
finite coupling $\Gamma \ne 0$. As shown in \figref{fig:smfig5}a, in the absence of coupling between molecule a single step appear in each molecule, which is associated to a single
localized mode in each site (\figref{fig:smfig5}b). In the presence of finite coupling between the molecules, a single steps transforms in a band of steps (\figref{fig:smfig5}c), which is associated to the emergence of propagating triplons. The manifold of steps is associated to the different many-body modes observed in the spin-structure factor (\figref{fig:smfig5}d), whose fine structure features are washed away by the smearing. In the absence of disorder (\figref{fig:smfig5}e,f), the internal structure of the
steps (\figref{fig:smfig5}e) would correspond to the confined triplon modes (\figref{fig:smfig5}f), that
would be also washed away by increasing smearing.

We finally comment on the disorder average in Fig.~5d in the main manuscript. We used open boundary conditions
for one-dimensional and two-dimensional geometries. One-dimensional geometries were taken with $L=N=10$,
and two-dimensional geometries were triangular lattice islands with $N=19$, comparable with the sizes found in the experiment. The open boundary conditions
make statistical distribution multi-modal. In the case of one-dimensional chains, its multimodal
nature stems from the different between edge and bulk sites. For two-dimensional islands, the
multimodal nature stems from the difference between edge sites, corner sites and bulk sites \suppcite{2022arXiv221207893K}.


\begin{thebibliography}{53}%
\makeatletter
\providecommand \@ifxundefined [1]{%
 \@ifx{#1\undefined}
}%
\providecommand \@ifnum [1]{%
 \ifnum #1\expandafter \@firstoftwo
 \else \expandafter \@secondoftwo
 \fi
}%
\providecommand \@ifx [1]{%
 \ifx #1\expandafter \@firstoftwo
 \else \expandafter \@secondoftwo
 \fi
}%
\providecommand \natexlab [1]{#1}%
\providecommand \enquote  [1]{``#1''}%
\providecommand \bibnamefont  [1]{#1}%
\providecommand \bibfnamefont [1]{#1}%
\providecommand \citenamefont [1]{#1}%
\providecommand \href@noop [0]{\@secondoftwo}%
\providecommand \href [0]{\begingroup \@sanitize@url \@href}%
\providecommand \@href[1]{\@@startlink{#1}\@@href}%
\providecommand \@@href[1]{\endgroup#1\@@endlink}%
\providecommand \@sanitize@url [0]{\catcode `\\12\catcode `\$12\catcode
  `\&12\catcode `\#12\catcode `\^12\catcode `\_12\catcode `\%12\relax}%
\providecommand \@@startlink[1]{}%
\providecommand \@@endlink[0]{}%
\providecommand \url  [0]{\begingroup\@sanitize@url \@url }%
\providecommand \@url [1]{\endgroup\@href {#1}{\urlprefix }}%
\providecommand \urlprefix  [0]{URL }%
\providecommand \Eprint [0]{\href }%
\providecommand \doibase [0]{https://doi.org/}%
\providecommand \selectlanguage [0]{\@gobble}%
\providecommand \bibinfo  [0]{\@secondoftwo}%
\providecommand \bibfield  [0]{\@secondoftwo}%
\providecommand \translation [1]{[#1]}%
\providecommand \BibitemOpen [0]{}%
\providecommand \bibitemStop [0]{}%
\providecommand \bibitemNoStop [0]{.\EOS\space}%
\providecommand \EOS [0]{\spacefactor3000\relax}%
\providecommand \BibitemShut  [1]{\csname bibitem#1\endcsname}%
\let\auto@bib@innerbib\@empty
\bibitem [{\citenamefont {Sachdev}(2008)}]{sachdev2008}%
  \BibitemOpen
  \bibfield  {author} {\bibinfo {author} {\bibfnamefont {S.}~\bibnamefont
  {Sachdev}},\ }\bibfield  {title} {\bibinfo {title} {Quantum magnetism and
  criticality},\ }\href {https://doi.org/10.1038/nphys894} {\bibfield
  {journal} {\bibinfo  {journal} {Nat. Phys.}\ }\textbf {\bibinfo {volume}
  {4}},\ \bibinfo {pages} {173} (\bibinfo {year} {2008})}\BibitemShut {NoStop}%
\bibitem [{\citenamefont {Vasiliev}\ \emph {et~al.}(2018)\citenamefont
  {Vasiliev}, \citenamefont {Volkova}, \citenamefont {Zvereva},\ and\
  \citenamefont {Markina}}]{vasiliev2018}%
  \BibitemOpen
  \bibfield  {author} {\bibinfo {author} {\bibfnamefont {A.}~\bibnamefont
  {Vasiliev}}, \bibinfo {author} {\bibfnamefont {O.}~\bibnamefont {Volkova}},
  \bibinfo {author} {\bibfnamefont {E.}~\bibnamefont {Zvereva}},\ and\ \bibinfo
  {author} {\bibfnamefont {M.}~\bibnamefont {Markina}},\ }\bibfield  {title}
  {\bibinfo {title} {Milestones of {low-D} quantum magnetism},\ }\href
  {https://doi.org/10.1038/s41535-018-0090-7} {\bibfield  {journal} {\bibinfo
  {journal} {npj Quantum Mater.}\ }\textbf {\bibinfo {volume} {3}},\ \bibinfo
  {pages} {18} (\bibinfo {year} {2018})}\BibitemShut {NoStop}%
\bibitem [{\citenamefont {Malki}\ and\ \citenamefont
  {Uhrig}(2020)}]{malki2020}%
  \BibitemOpen
  \bibfield  {author} {\bibinfo {author} {\bibfnamefont {M.}~\bibnamefont
  {Malki}}\ and\ \bibinfo {author} {\bibfnamefont {G.~S.}\ \bibnamefont
  {Uhrig}},\ }\bibfield  {title} {\bibinfo {title} {Topological magnetic
  excitations},\ }\href {https://doi.org/10.1209/0295-5075/132/20003}
  {\bibfield  {journal} {\bibinfo  {journal} {EPL}\ }\textbf {\bibinfo {volume}
  {132}},\ \bibinfo {pages} {20003} (\bibinfo {year} {2020})}\BibitemShut
  {NoStop}%
\bibitem [{\citenamefont {Chumak}\ \emph {et~al.}(2015)\citenamefont {Chumak},
  \citenamefont {Vasyuchka}, \citenamefont {Serga},\ and\ \citenamefont
  {Hillebrands}}]{chumak2015}%
  \BibitemOpen
  \bibfield  {author} {\bibinfo {author} {\bibfnamefont {A.~V.}\ \bibnamefont
  {Chumak}}, \bibinfo {author} {\bibfnamefont {V.~I.}\ \bibnamefont
  {Vasyuchka}}, \bibinfo {author} {\bibfnamefont {A.~A.}\ \bibnamefont
  {Serga}},\ and\ \bibinfo {author} {\bibfnamefont {B.}~\bibnamefont
  {Hillebrands}},\ }\bibfield  {title} {\bibinfo {title} {Magnon spintronics},\
  }\href {https://doi.org/10.1038/nphys3347} {\bibfield  {journal} {\bibinfo
  {journal} {Nat. Phys.}\ }\textbf {\bibinfo {volume} {11}},\ \bibinfo {pages}
  {453} (\bibinfo {year} {2015})}\BibitemShut {NoStop}%
\bibitem [{\citenamefont {Faddeev}\ and\ \citenamefont
  {Takhtajan}(1981)}]{faddeev1981}%
  \BibitemOpen
  \bibfield  {author} {\bibinfo {author} {\bibfnamefont {L.}~\bibnamefont
  {Faddeev}}\ and\ \bibinfo {author} {\bibfnamefont {L.}~\bibnamefont
  {Takhtajan}},\ }\bibfield  {title} {\bibinfo {title} {What is the spin of a
  spin wave?},\ }\href {https://doi.org/10.1016/0375-9601(81)90335-2}
  {\bibfield  {journal} {\bibinfo  {journal} {Phys. Lett. A}\ }\textbf
  {\bibinfo {volume} {85}},\ \bibinfo {pages} {375} (\bibinfo {year}
  {1981})}\BibitemShut {NoStop}%
\bibitem [{\citenamefont {Kukushkin}\ \emph {et~al.}(2009)\citenamefont
  {Kukushkin}, \citenamefont {Smet}, \citenamefont {Scarola}, \citenamefont
  {Umansky},\ and\ \citenamefont {von Klitzing}}]{kukushkin2009}%
  \BibitemOpen
  \bibfield  {author} {\bibinfo {author} {\bibfnamefont {I.~V.}\ \bibnamefont
  {Kukushkin}}, \bibinfo {author} {\bibfnamefont {J.~H.}\ \bibnamefont {Smet}},
  \bibinfo {author} {\bibfnamefont {V.~W.}\ \bibnamefont {Scarola}}, \bibinfo
  {author} {\bibfnamefont {V.}~\bibnamefont {Umansky}},\ and\ \bibinfo {author}
  {\bibfnamefont {K.}~\bibnamefont {von Klitzing}},\ }\bibfield  {title}
  {\bibinfo {title} {Dispersion of the excitations of fractional quantum {Hall}
  states},\ }\href {https://doi.org/10.1126/science.1171472} {\bibfield
  {journal} {\bibinfo  {journal} {Science}\ }\textbf {\bibinfo {volume}
  {324}},\ \bibinfo {pages} {1044} (\bibinfo {year} {2009})}\BibitemShut
  {NoStop}%
\bibitem [{\citenamefont {Hao}\ and\ \citenamefont
  {Tchernyshyov}(2009)}]{hao2009}%
  \BibitemOpen
  \bibfield  {author} {\bibinfo {author} {\bibfnamefont {Z.}~\bibnamefont
  {Hao}}\ and\ \bibinfo {author} {\bibfnamefont {O.}~\bibnamefont
  {Tchernyshyov}},\ }\bibfield  {title} {\bibinfo {title} {Fermionic spin
  excitations in two- and three-dimensional antiferromagnets},\ }\href
  {https://doi.org/10.1103/PhysRevLett.103.187203} {\bibfield  {journal}
  {\bibinfo  {journal} {Phys. Rev. Lett.}\ }\textbf {\bibinfo {volume} {103}},\
  \bibinfo {pages} {187203} (\bibinfo {year} {2009})}\BibitemShut {NoStop}%
\bibitem [{\citenamefont {Han}\ \emph {et~al.}(2012)\citenamefont {Han},
  \citenamefont {Helton}, \citenamefont {Chu}, \citenamefont {Nocera},
  \citenamefont {Rodriguez-Rivera}, \citenamefont {Broholm},\ and\
  \citenamefont {Lee}}]{han2012}%
  \BibitemOpen
  \bibfield  {author} {\bibinfo {author} {\bibfnamefont {T.-H.}\ \bibnamefont
  {Han}}, \bibinfo {author} {\bibfnamefont {J.~S.}\ \bibnamefont {Helton}},
  \bibinfo {author} {\bibfnamefont {S.}~\bibnamefont {Chu}}, \bibinfo {author}
  {\bibfnamefont {D.~G.}\ \bibnamefont {Nocera}}, \bibinfo {author}
  {\bibfnamefont {J.~A.}\ \bibnamefont {Rodriguez-Rivera}}, \bibinfo {author}
  {\bibfnamefont {C.}~\bibnamefont {Broholm}},\ and\ \bibinfo {author}
  {\bibfnamefont {Y.~S.}\ \bibnamefont {Lee}},\ }\bibfield  {title} {\bibinfo
  {title} {Fractionalized excitations in the spin-liquid state of a
  kagome-lattice antiferromagnet},\ }\href
  {https://doi.org/10.1038/nature11659} {\bibfield  {journal} {\bibinfo
  {journal} {Nature}\ }\textbf {\bibinfo {volume} {492}},\ \bibinfo {pages}
  {406} (\bibinfo {year} {2012})}\BibitemShut {NoStop}%
\bibitem [{\citenamefont {Mourigal}\ \emph {et~al.}(2013)\citenamefont
  {Mourigal}, \citenamefont {Enderle}, \citenamefont {Kl\"{o}pperpieper},
  \citenamefont {Caux}, \citenamefont {Stunault},\ and\ \citenamefont
  {R{\o}nnow}}]{mourigal2013}%
  \BibitemOpen
  \bibfield  {author} {\bibinfo {author} {\bibfnamefont {M.}~\bibnamefont
  {Mourigal}}, \bibinfo {author} {\bibfnamefont {M.}~\bibnamefont {Enderle}},
  \bibinfo {author} {\bibfnamefont {A.}~\bibnamefont {Kl\"{o}pperpieper}},
  \bibinfo {author} {\bibfnamefont {J.-S.}\ \bibnamefont {Caux}}, \bibinfo
  {author} {\bibfnamefont {A.}~\bibnamefont {Stunault}},\ and\ \bibinfo
  {author} {\bibfnamefont {H.~M.}\ \bibnamefont {R{\o}nnow}},\ }\bibfield
  {title} {\bibinfo {title} {Fractional spinon excitations in the quantum
  {Heisenberg} antiferromagnetic chain},\ }\href
  {https://doi.org/10.1038/nphys2652} {\bibfield  {journal} {\bibinfo
  {journal} {Nat. Phys.}\ }\textbf {\bibinfo {volume} {9}},\ \bibinfo {pages}
  {435} (\bibinfo {year} {2013})}\BibitemShut {NoStop}%
\bibitem [{\citenamefont {Piazza}\ \emph {et~al.}(2014)\citenamefont {Piazza},
  \citenamefont {Mourigal}, \citenamefont {Christensen}, \citenamefont
  {Nilsen}, \citenamefont {Tregenna-Piggott}, \citenamefont {Perring},
  \citenamefont {Enderle}, \citenamefont {McMorrow}, \citenamefont {Ivanov},\
  and\ \citenamefont {R{\o}nnow}}]{dalla2015}%
  \BibitemOpen
  \bibfield  {author} {\bibinfo {author} {\bibfnamefont {B.~D.}\ \bibnamefont
  {Piazza}}, \bibinfo {author} {\bibfnamefont {M.}~\bibnamefont {Mourigal}},
  \bibinfo {author} {\bibfnamefont {N.~B.}\ \bibnamefont {Christensen}},
  \bibinfo {author} {\bibfnamefont {G.~J.}\ \bibnamefont {Nilsen}}, \bibinfo
  {author} {\bibfnamefont {P.}~\bibnamefont {Tregenna-Piggott}}, \bibinfo
  {author} {\bibfnamefont {T.~G.}\ \bibnamefont {Perring}}, \bibinfo {author}
  {\bibfnamefont {M.}~\bibnamefont {Enderle}}, \bibinfo {author} {\bibfnamefont
  {D.~F.}\ \bibnamefont {McMorrow}}, \bibinfo {author} {\bibfnamefont {D.~A.}\
  \bibnamefont {Ivanov}},\ and\ \bibinfo {author} {\bibfnamefont {H.~M.}\
  \bibnamefont {R{\o}nnow}},\ }\bibfield  {title} {\bibinfo {title} {Fractional
  excitations in the square-lattice quantum antiferromagnet},\ }\href
  {https://doi.org/10.1038/nphys3172} {\bibfield  {journal} {\bibinfo
  {journal} {Nat. Phys.}\ }\textbf {\bibinfo {volume} {11}},\ \bibinfo {pages}
  {62} (\bibinfo {year} {2014})}\BibitemShut {NoStop}%
\bibitem [{\citenamefont {Sachdev}\ and\ \citenamefont
  {Bhatt}(1990)}]{PhysRevB.41.9323}%
  \BibitemOpen
  \bibfield  {author} {\bibinfo {author} {\bibfnamefont {S.}~\bibnamefont
  {Sachdev}}\ and\ \bibinfo {author} {\bibfnamefont {R.~N.}\ \bibnamefont
  {Bhatt}},\ }\bibfield  {title} {\bibinfo {title} {Bond-operator
  representation of quantum spins: Mean-field theory of frustrated quantum
  {Heisenberg} antiferromagnets},\ }\href
  {https://doi.org/10.1103/PhysRevB.41.9323} {\bibfield  {journal} {\bibinfo
  {journal} {Phys. Rev. B}\ }\textbf {\bibinfo {volume} {41}},\ \bibinfo
  {pages} {9323} (\bibinfo {year} {1990})}\BibitemShut {NoStop}%
\bibitem [{\citenamefont {Cavadini}\ \emph {et~al.}(2001)\citenamefont
  {Cavadini}, \citenamefont {Heigold}, \citenamefont {Henggeler}, \citenamefont
  {Furrer}, \citenamefont {G\"udel}, \citenamefont {Kr\"amer},\ and\
  \citenamefont {Mutka}}]{cavadini2001}%
  \BibitemOpen
  \bibfield  {author} {\bibinfo {author} {\bibfnamefont {N.}~\bibnamefont
  {Cavadini}}, \bibinfo {author} {\bibfnamefont {G.}~\bibnamefont {Heigold}},
  \bibinfo {author} {\bibfnamefont {W.}~\bibnamefont {Henggeler}}, \bibinfo
  {author} {\bibfnamefont {A.}~\bibnamefont {Furrer}}, \bibinfo {author}
  {\bibfnamefont {H.-U.}\ \bibnamefont {G\"udel}}, \bibinfo {author}
  {\bibfnamefont {K.}~\bibnamefont {Kr\"amer}},\ and\ \bibinfo {author}
  {\bibfnamefont {H.}~\bibnamefont {Mutka}},\ }\bibfield  {title} {\bibinfo
  {title} {Magnetic excitations in the quantum spin system {TlCuCl}$_{3}$},\
  }\href {https://doi.org/10.1103/PhysRevB.63.172414} {\bibfield  {journal}
  {\bibinfo  {journal} {Phys. Rev. B}\ }\textbf {\bibinfo {volume} {63}},\
  \bibinfo {pages} {172414} (\bibinfo {year} {2001})}\BibitemShut {NoStop}%
\bibitem [{\citenamefont {Xu}\ \emph {et~al.}(2007)\citenamefont {Xu},
  \citenamefont {Broholm}, \citenamefont {Soh}, \citenamefont {Aeppli},
  \citenamefont {DiTusa}, \citenamefont {Chen}, \citenamefont {Kenzelmann},
  \citenamefont {Frost}, \citenamefont {Ito}, \citenamefont {Oka},\ and\
  \citenamefont {Takagi}}]{xu2007}%
  \BibitemOpen
  \bibfield  {author} {\bibinfo {author} {\bibfnamefont {G.}~\bibnamefont
  {Xu}}, \bibinfo {author} {\bibfnamefont {C.}~\bibnamefont {Broholm}},
  \bibinfo {author} {\bibfnamefont {Y.-A.}\ \bibnamefont {Soh}}, \bibinfo
  {author} {\bibfnamefont {G.}~\bibnamefont {Aeppli}}, \bibinfo {author}
  {\bibfnamefont {J.~F.}\ \bibnamefont {DiTusa}}, \bibinfo {author}
  {\bibfnamefont {Y.}~\bibnamefont {Chen}}, \bibinfo {author} {\bibfnamefont
  {M.}~\bibnamefont {Kenzelmann}}, \bibinfo {author} {\bibfnamefont {C.~D.}\
  \bibnamefont {Frost}}, \bibinfo {author} {\bibfnamefont {T.}~\bibnamefont
  {Ito}}, \bibinfo {author} {\bibfnamefont {K.}~\bibnamefont {Oka}},\ and\
  \bibinfo {author} {\bibfnamefont {H.}~\bibnamefont {Takagi}},\ }\bibfield
  {title} {\bibinfo {title} {Mesoscopic phase coherence in a quantum spin
  fluid},\ }\href {https://doi.org/10.1126/science.1143831} {\bibfield
  {journal} {\bibinfo  {journal} {Science}\ }\textbf {\bibinfo {volume}
  {317}},\ \bibinfo {pages} {1049} (\bibinfo {year} {2007})}\BibitemShut
  {NoStop}%
\bibitem [{\citenamefont {McClarty}\ \emph {et~al.}(2017)\citenamefont
  {McClarty}, \citenamefont {Kr\"{u}ger}, \citenamefont {Guidi}, \citenamefont
  {Parker}, \citenamefont {Refson}, \citenamefont {Parker}, \citenamefont
  {Prabhakaran},\ and\ \citenamefont {Coldea}}]{McClarty2017}%
  \BibitemOpen
  \bibfield  {author} {\bibinfo {author} {\bibfnamefont {P.~A.}\ \bibnamefont
  {McClarty}}, \bibinfo {author} {\bibfnamefont {F.}~\bibnamefont
  {Kr\"{u}ger}}, \bibinfo {author} {\bibfnamefont {T.}~\bibnamefont {Guidi}},
  \bibinfo {author} {\bibfnamefont {S.~F.}\ \bibnamefont {Parker}}, \bibinfo
  {author} {\bibfnamefont {K.}~\bibnamefont {Refson}}, \bibinfo {author}
  {\bibfnamefont {A.~W.}\ \bibnamefont {Parker}}, \bibinfo {author}
  {\bibfnamefont {D.}~\bibnamefont {Prabhakaran}},\ and\ \bibinfo {author}
  {\bibfnamefont {R.}~\bibnamefont {Coldea}},\ }\bibfield  {title} {\bibinfo
  {title} {Topological triplon modes and bound states in a
  {Shastry{\textendash}Sutherland} magnet},\ }\href
  {https://doi.org/10.1038/nphys4117} {\bibfield  {journal} {\bibinfo
  {journal} {Nat. Phys.}\ }\textbf {\bibinfo {volume} {13}},\ \bibinfo {pages}
  {736} (\bibinfo {year} {2017})}\BibitemShut {NoStop}%
\bibitem [{\citenamefont {Notbohm}\ \emph {et~al.}(2007)\citenamefont
  {Notbohm}, \citenamefont {Ribeiro}, \citenamefont {Lake}, \citenamefont
  {Tennant}, \citenamefont {Schmidt}, \citenamefont {Uhrig}, \citenamefont
  {Hess}, \citenamefont {Klingeler}, \citenamefont {Behr}, \citenamefont
  {B\"uchner}, \citenamefont {Reehuis}, \citenamefont {Bewley}, \citenamefont
  {Frost}, \citenamefont {Manuel},\ and\ \citenamefont
  {Eccleston}}]{PhysRevLett.98.027403}%
  \BibitemOpen
  \bibfield  {author} {\bibinfo {author} {\bibfnamefont {S.}~\bibnamefont
  {Notbohm}}, \bibinfo {author} {\bibfnamefont {P.}~\bibnamefont {Ribeiro}},
  \bibinfo {author} {\bibfnamefont {B.}~\bibnamefont {Lake}}, \bibinfo {author}
  {\bibfnamefont {D.~A.}\ \bibnamefont {Tennant}}, \bibinfo {author}
  {\bibfnamefont {K.~P.}\ \bibnamefont {Schmidt}}, \bibinfo {author}
  {\bibfnamefont {G.~S.}\ \bibnamefont {Uhrig}}, \bibinfo {author}
  {\bibfnamefont {C.}~\bibnamefont {Hess}}, \bibinfo {author} {\bibfnamefont
  {R.}~\bibnamefont {Klingeler}}, \bibinfo {author} {\bibfnamefont
  {G.}~\bibnamefont {Behr}}, \bibinfo {author} {\bibfnamefont {B.}~\bibnamefont
  {B\"uchner}}, \bibinfo {author} {\bibfnamefont {M.}~\bibnamefont {Reehuis}},
  \bibinfo {author} {\bibfnamefont {R.~I.}\ \bibnamefont {Bewley}}, \bibinfo
  {author} {\bibfnamefont {C.~D.}\ \bibnamefont {Frost}}, \bibinfo {author}
  {\bibfnamefont {P.}~\bibnamefont {Manuel}},\ and\ \bibinfo {author}
  {\bibfnamefont {R.~S.}\ \bibnamefont {Eccleston}},\ }\bibfield  {title}
  {\bibinfo {title} {One- and two-triplon spectra of a cuprate ladder},\ }\href
  {https://doi.org/10.1103/PhysRevLett.98.027403} {\bibfield  {journal}
  {\bibinfo  {journal} {Phys. Rev. Lett.}\ }\textbf {\bibinfo {volume} {98}},\
  \bibinfo {pages} {027403} (\bibinfo {year} {2007})}\BibitemShut {NoStop}%
\bibitem [{\citenamefont {Nawa}\ \emph {et~al.}(2019)\citenamefont {Nawa},
  \citenamefont {Tanaka}, \citenamefont {Kurita}, \citenamefont {Sato},
  \citenamefont {Sugiyama}, \citenamefont {Uekusa}, \citenamefont
  {Ohira-Kawamura}, \citenamefont {Nakajima},\ and\ \citenamefont
  {Tanaka}}]{Nawa2019}%
  \BibitemOpen
  \bibfield  {author} {\bibinfo {author} {\bibfnamefont {K.}~\bibnamefont
  {Nawa}}, \bibinfo {author} {\bibfnamefont {K.}~\bibnamefont {Tanaka}},
  \bibinfo {author} {\bibfnamefont {N.}~\bibnamefont {Kurita}}, \bibinfo
  {author} {\bibfnamefont {T.~J.}\ \bibnamefont {Sato}}, \bibinfo {author}
  {\bibfnamefont {H.}~\bibnamefont {Sugiyama}}, \bibinfo {author}
  {\bibfnamefont {H.}~\bibnamefont {Uekusa}}, \bibinfo {author} {\bibfnamefont
  {S.}~\bibnamefont {Ohira-Kawamura}}, \bibinfo {author} {\bibfnamefont
  {K.}~\bibnamefont {Nakajima}},\ and\ \bibinfo {author} {\bibfnamefont
  {H.}~\bibnamefont {Tanaka}},\ }\bibfield  {title} {\bibinfo {title} {Triplon
  band splitting and topologically protected edge states in the dimerized
  antiferromagnet},\ }\href {https://doi.org/10.1038/s41467-019-10091-6}
  {\bibfield  {journal} {\bibinfo  {journal} {Nat. Commun.}\ }\textbf {\bibinfo
  {volume} {10}},\ \bibinfo {pages} {2096} (\bibinfo {year}
  {2019})}\BibitemShut {NoStop}%
\bibitem [{\citenamefont {Coey}(2010)}]{coey2010magnetism}%
  \BibitemOpen
  \bibfield  {author} {\bibinfo {author} {\bibfnamefont {J.~M.}\ \bibnamefont
  {Coey}},\ }\href@noop {} {\emph {\bibinfo {title} {Magnetism and magnetic
  materials}}}\ (\bibinfo  {publisher} {Cambridge University Press},\ \bibinfo
  {year} {2010})\BibitemShut {NoStop}%
\bibitem [{\citenamefont {Coronado}(2019)}]{Coronado2019}%
  \BibitemOpen
  \bibfield  {author} {\bibinfo {author} {\bibfnamefont {E.}~\bibnamefont
  {Coronado}},\ }\bibfield  {title} {\bibinfo {title} {Molecular magnetism:
  from chemical design to spin control in molecules, materials and devices},\
  }\href {https://doi.org/10.1038/s41578-019-0146-8} {\bibfield  {journal}
  {\bibinfo  {journal} {Nat. Rev. Mater.}\ }\textbf {\bibinfo {volume} {5}},\
  \bibinfo {pages} {87} (\bibinfo {year} {2019})}\BibitemShut {NoStop}%
\bibitem [{\citenamefont {Kohno}\ \emph {et~al.}(2007)\citenamefont {Kohno},
  \citenamefont {Starykh},\ and\ \citenamefont {Balents}}]{Kohno2007}%
  \BibitemOpen
  \bibfield  {author} {\bibinfo {author} {\bibfnamefont {M.}~\bibnamefont
  {Kohno}}, \bibinfo {author} {\bibfnamefont {O.~A.}\ \bibnamefont {Starykh}},\
  and\ \bibinfo {author} {\bibfnamefont {L.}~\bibnamefont {Balents}},\
  }\bibfield  {title} {\bibinfo {title} {Spinons and triplons in spatially
  anisotropic frustrated antiferromagnets},\ }\href
  {https://doi.org/10.1038/nphys749} {\bibfield  {journal} {\bibinfo  {journal}
  {Nat. Phys.}\ }\textbf {\bibinfo {volume} {3}},\ \bibinfo {pages} {790}
  (\bibinfo {year} {2007})}\BibitemShut {NoStop}%
\bibitem [{\citenamefont {R\"uegg}\ \emph {et~al.}(2008)\citenamefont
  {R\"uegg}, \citenamefont {Normand}, \citenamefont {Matsumoto}, \citenamefont
  {Furrer}, \citenamefont {McMorrow}, \citenamefont {Kr\"amer}, \citenamefont
  {G\"udel}, \citenamefont {Gvasaliya}, \citenamefont {Mutka},\ and\
  \citenamefont {Boehm}}]{PhysRevLett.100.205701}%
  \BibitemOpen
  \bibfield  {author} {\bibinfo {author} {\bibfnamefont {C.}~\bibnamefont
  {R\"uegg}}, \bibinfo {author} {\bibfnamefont {B.}~\bibnamefont {Normand}},
  \bibinfo {author} {\bibfnamefont {M.}~\bibnamefont {Matsumoto}}, \bibinfo
  {author} {\bibfnamefont {A.}~\bibnamefont {Furrer}}, \bibinfo {author}
  {\bibfnamefont {D.~F.}\ \bibnamefont {McMorrow}}, \bibinfo {author}
  {\bibfnamefont {K.~W.}\ \bibnamefont {Kr\"amer}}, \bibinfo {author}
  {\bibfnamefont {H.~U.}\ \bibnamefont {G\"udel}}, \bibinfo {author}
  {\bibfnamefont {S.~N.}\ \bibnamefont {Gvasaliya}}, \bibinfo {author}
  {\bibfnamefont {H.}~\bibnamefont {Mutka}},\ and\ \bibinfo {author}
  {\bibfnamefont {M.}~\bibnamefont {Boehm}},\ }\bibfield  {title} {\bibinfo
  {title} {Quantum magnets under pressure: Controlling elementary excitations
  in {TlCuCl}$_{3}$},\ }\href {https://doi.org/10.1103/PhysRevLett.100.205701}
  {\bibfield  {journal} {\bibinfo  {journal} {Phys. Rev. Lett.}\ }\textbf
  {\bibinfo {volume} {100}},\ \bibinfo {pages} {205701} (\bibinfo {year}
  {2008})}\BibitemShut {NoStop}%
\bibitem [{\citenamefont {Schlappa}\ \emph {et~al.}(2009)\citenamefont
  {Schlappa}, \citenamefont {Schmitt}, \citenamefont {Vernay}, \citenamefont
  {Strocov}, \citenamefont {Ilakovac}, \citenamefont {Thielemann},
  \citenamefont {R\o{}nnow}, \citenamefont {Vanishri}, \citenamefont
  {Piazzalunga}, \citenamefont {Wang}, \citenamefont {Braicovich},
  \citenamefont {Ghiringhelli}, \citenamefont {Marin}, \citenamefont {Mesot},
  \citenamefont {Delley},\ and\ \citenamefont
  {Patthey}}]{PhysRevLett.103.047401}%
  \BibitemOpen
  \bibfield  {author} {\bibinfo {author} {\bibfnamefont {J.}~\bibnamefont
  {Schlappa}}, \bibinfo {author} {\bibfnamefont {T.}~\bibnamefont {Schmitt}},
  \bibinfo {author} {\bibfnamefont {F.}~\bibnamefont {Vernay}}, \bibinfo
  {author} {\bibfnamefont {V.~N.}\ \bibnamefont {Strocov}}, \bibinfo {author}
  {\bibfnamefont {V.}~\bibnamefont {Ilakovac}}, \bibinfo {author}
  {\bibfnamefont {B.}~\bibnamefont {Thielemann}}, \bibinfo {author}
  {\bibfnamefont {H.~M.}\ \bibnamefont {R\o{}nnow}}, \bibinfo {author}
  {\bibfnamefont {S.}~\bibnamefont {Vanishri}}, \bibinfo {author}
  {\bibfnamefont {A.}~\bibnamefont {Piazzalunga}}, \bibinfo {author}
  {\bibfnamefont {X.}~\bibnamefont {Wang}}, \bibinfo {author} {\bibfnamefont
  {L.}~\bibnamefont {Braicovich}}, \bibinfo {author} {\bibfnamefont
  {G.}~\bibnamefont {Ghiringhelli}}, \bibinfo {author} {\bibfnamefont
  {C.}~\bibnamefont {Marin}}, \bibinfo {author} {\bibfnamefont
  {J.}~\bibnamefont {Mesot}}, \bibinfo {author} {\bibfnamefont
  {B.}~\bibnamefont {Delley}},\ and\ \bibinfo {author} {\bibfnamefont
  {L.}~\bibnamefont {Patthey}},\ }\bibfield  {title} {\bibinfo {title}
  {Collective magnetic excitations in the spin ladder
  {Sr$_{14}$Cu$_{24}$O$_{41}$} measured using high-resolution resonant
  inelastic x-ray scattering},\ }\href
  {https://doi.org/10.1103/PhysRevLett.103.047401} {\bibfield  {journal}
  {\bibinfo  {journal} {Phys. Rev. Lett.}\ }\textbf {\bibinfo {volume} {103}},\
  \bibinfo {pages} {047401} (\bibinfo {year} {2009})}\BibitemShut {NoStop}%
\bibitem [{\citenamefont {Zhitomirsky}\ and\ \citenamefont
  {Chernyshev}(2013)}]{RevModPhys.85.219}%
  \BibitemOpen
  \bibfield  {author} {\bibinfo {author} {\bibfnamefont {M.~E.}\ \bibnamefont
  {Zhitomirsky}}\ and\ \bibinfo {author} {\bibfnamefont {A.~L.}\ \bibnamefont
  {Chernyshev}},\ }\bibfield  {title} {\bibinfo {title} {Colloquium:
  Spontaneous magnon decays},\ }\href
  {https://doi.org/10.1103/RevModPhys.85.219} {\bibfield  {journal} {\bibinfo
  {journal} {Rev. Mod. Phys.}\ }\textbf {\bibinfo {volume} {85}},\ \bibinfo
  {pages} {219} (\bibinfo {year} {2013})}\BibitemShut {NoStop}%
\bibitem [{\citenamefont {Zayed}\ \emph {et~al.}(2014)\citenamefont {Zayed},
  \citenamefont {R\"uegg}, \citenamefont {Str\"assle}, \citenamefont {Stuhr},
  \citenamefont {Roessli}, \citenamefont {Ay}, \citenamefont {Mesot},
  \citenamefont {Link}, \citenamefont {Pomjakushina}, \citenamefont
  {Stingaciu}, \citenamefont {Conder},\ and\ \citenamefont
  {R\o{}nnow}}]{PhysRevLett.113.067201}%
  \BibitemOpen
  \bibfield  {author} {\bibinfo {author} {\bibfnamefont {M.~E.}\ \bibnamefont
  {Zayed}}, \bibinfo {author} {\bibfnamefont {C.}~\bibnamefont {R\"uegg}},
  \bibinfo {author} {\bibfnamefont {T.}~\bibnamefont {Str\"assle}}, \bibinfo
  {author} {\bibfnamefont {U.}~\bibnamefont {Stuhr}}, \bibinfo {author}
  {\bibfnamefont {B.}~\bibnamefont {Roessli}}, \bibinfo {author} {\bibfnamefont
  {M.}~\bibnamefont {Ay}}, \bibinfo {author} {\bibfnamefont {J.}~\bibnamefont
  {Mesot}}, \bibinfo {author} {\bibfnamefont {P.}~\bibnamefont {Link}},
  \bibinfo {author} {\bibfnamefont {E.}~\bibnamefont {Pomjakushina}}, \bibinfo
  {author} {\bibfnamefont {M.}~\bibnamefont {Stingaciu}}, \bibinfo {author}
  {\bibfnamefont {K.}~\bibnamefont {Conder}},\ and\ \bibinfo {author}
  {\bibfnamefont {H.~M.}\ \bibnamefont {R\o{}nnow}},\ }\bibfield  {title}
  {\bibinfo {title} {Correlated decay of triplet excitations in the
  {Shastry-Sutherland} compound {SrCu$_{2}$(BO$_{3}$)$_{2}$}},\ }\href
  {https://doi.org/10.1103/PhysRevLett.113.067201} {\bibfield  {journal}
  {\bibinfo  {journal} {Phys. Rev. Lett.}\ }\textbf {\bibinfo {volume} {113}},\
  \bibinfo {pages} {067201} (\bibinfo {year} {2014})}\BibitemShut {NoStop}%
\bibitem [{\citenamefont {Franke}\ \emph {et~al.}(2011)\citenamefont {Franke},
  \citenamefont {Schulze},\ and\ \citenamefont {Pascual}}]{Franke2011}%
  \BibitemOpen
  \bibfield  {author} {\bibinfo {author} {\bibfnamefont {K.~J.}\ \bibnamefont
  {Franke}}, \bibinfo {author} {\bibfnamefont {G.}~\bibnamefont {Schulze}},\
  and\ \bibinfo {author} {\bibfnamefont {J.~I.}\ \bibnamefont {Pascual}},\
  }\bibfield  {title} {\bibinfo {title} {Competition of superconducting
  phenomena and {Kondo} screening at the nanoscale},\ }\href
  {https://doi.org/10.1126/science.1202204} {\bibfield  {journal} {\bibinfo
  {journal} {Science}\ }\textbf {\bibinfo {volume} {332}},\ \bibinfo {pages}
  {940} (\bibinfo {year} {2011})}\BibitemShut {NoStop}%
\bibitem [{\citenamefont {Malavolti}\ \emph {et~al.}(2018)\citenamefont
  {Malavolti}, \citenamefont {Briganti}, \citenamefont {H\"{a}nze},
  \citenamefont {Serrano}, \citenamefont {Cimatti}, \citenamefont {McMurtrie},
  \citenamefont {Otero}, \citenamefont {Ohresser}, \citenamefont {Totti},
  \citenamefont {Mannini}, \citenamefont {Sessoli},\ and\ \citenamefont
  {Loth}}]{malavolti2018}%
  \BibitemOpen
  \bibfield  {author} {\bibinfo {author} {\bibfnamefont {L.}~\bibnamefont
  {Malavolti}}, \bibinfo {author} {\bibfnamefont {M.}~\bibnamefont {Briganti}},
  \bibinfo {author} {\bibfnamefont {M.}~\bibnamefont {H\"{a}nze}}, \bibinfo
  {author} {\bibfnamefont {G.}~\bibnamefont {Serrano}}, \bibinfo {author}
  {\bibfnamefont {I.}~\bibnamefont {Cimatti}}, \bibinfo {author} {\bibfnamefont
  {G.}~\bibnamefont {McMurtrie}}, \bibinfo {author} {\bibfnamefont
  {E.}~\bibnamefont {Otero}}, \bibinfo {author} {\bibfnamefont
  {P.}~\bibnamefont {Ohresser}}, \bibinfo {author} {\bibfnamefont
  {F.}~\bibnamefont {Totti}}, \bibinfo {author} {\bibfnamefont
  {M.}~\bibnamefont {Mannini}}, \bibinfo {author} {\bibfnamefont
  {R.}~\bibnamefont {Sessoli}},\ and\ \bibinfo {author} {\bibfnamefont
  {S.}~\bibnamefont {Loth}},\ }\bibfield  {title} {\bibinfo {title} {Tunable
  spin{\textendash}superconductor coupling of spin 1/2 vanadyl phthalocyanine
  molecules},\ }\href {https://doi.org/10.1021/acs.nanolett.8b03921} {\bibfield
   {journal} {\bibinfo  {journal} {Nano Lett.}\ }\textbf {\bibinfo {volume}
  {18}},\ \bibinfo {pages} {7955} (\bibinfo {year} {2018})}\BibitemShut
  {NoStop}%
\bibitem [{\citenamefont {Kezilebieke}\ \emph {et~al.}(2018)\citenamefont
  {Kezilebieke}, \citenamefont {Dvorak}, \citenamefont {Ojanen},\ and\
  \citenamefont {Liljeroth}}]{Kezilebieke2018}%
  \BibitemOpen
  \bibfield  {author} {\bibinfo {author} {\bibfnamefont {S.}~\bibnamefont
  {Kezilebieke}}, \bibinfo {author} {\bibfnamefont {M.}~\bibnamefont {Dvorak}},
  \bibinfo {author} {\bibfnamefont {T.}~\bibnamefont {Ojanen}},\ and\ \bibinfo
  {author} {\bibfnamefont {P.}~\bibnamefont {Liljeroth}},\ }\bibfield  {title}
  {\bibinfo {title} {Coupled {Yu-Shiba-Rusinov} states in molecular dimers on
  {NbSe}$_2$},\ }\href {https://doi.org/10.1021/acs.nanolett.7b05050}
  {\bibfield  {journal} {\bibinfo  {journal} {Nano Lett.}\ }\textbf {\bibinfo
  {volume} {18}},\ \bibinfo {pages} {2311} (\bibinfo {year}
  {2018})}\BibitemShut {NoStop}%
\bibitem [{\citenamefont {Kezilebieke}\ \emph {et~al.}(2019)\citenamefont
  {Kezilebieke}, \citenamefont {{\v{Z}}itko}, \citenamefont {Dvorak},
  \citenamefont {Ojanen},\ and\ \citenamefont {Liljeroth}}]{Kezilebieke2019}%
  \BibitemOpen
  \bibfield  {author} {\bibinfo {author} {\bibfnamefont {S.}~\bibnamefont
  {Kezilebieke}}, \bibinfo {author} {\bibfnamefont {R.}~\bibnamefont
  {{\v{Z}}itko}}, \bibinfo {author} {\bibfnamefont {M.}~\bibnamefont {Dvorak}},
  \bibinfo {author} {\bibfnamefont {T.}~\bibnamefont {Ojanen}},\ and\ \bibinfo
  {author} {\bibfnamefont {P.}~\bibnamefont {Liljeroth}},\ }\bibfield  {title}
  {\bibinfo {title} {Observation of coexistence of {Yu-Shiba-Rusinov} states
  and spin-flip excitations},\ }\href
  {https://doi.org/10.1021/acs.nanolett.9b01583} {\bibfield  {journal}
  {\bibinfo  {journal} {Nano Lett.}\ }\textbf {\bibinfo {volume} {19}},\
  \bibinfo {pages} {4614} (\bibinfo {year} {2019})}\BibitemShut {NoStop}%
\bibitem [{\citenamefont {Wang}\ \emph {et~al.}(2021)\citenamefont {Wang},
  \citenamefont {Arabi}, \citenamefont {Kern},\ and\ \citenamefont
  {Ternes}}]{Wang2021}%
  \BibitemOpen
  \bibfield  {author} {\bibinfo {author} {\bibfnamefont {Y.}~\bibnamefont
  {Wang}}, \bibinfo {author} {\bibfnamefont {S.}~\bibnamefont {Arabi}},
  \bibinfo {author} {\bibfnamefont {K.}~\bibnamefont {Kern}},\ and\ \bibinfo
  {author} {\bibfnamefont {M.}~\bibnamefont {Ternes}},\ }\bibfield  {title}
  {\bibinfo {title} {Symmetry mediated tunable molecular magnetism on a {2D}
  material},\ }\href {https://doi.org/10.1038/s42005-021-00601-8} {\bibfield
  {journal} {\bibinfo  {journal} {Commun. Phys.}\ }\textbf {\bibinfo {volume}
  {4}},\ \bibinfo {pages} {103} (\bibinfo {year} {2021})}\BibitemShut {NoStop}%
\bibitem [{\citenamefont {Ternes}(2015)}]{ternes2015}%
  \BibitemOpen
  \bibfield  {author} {\bibinfo {author} {\bibfnamefont {M.}~\bibnamefont
  {Ternes}},\ }\bibfield  {title} {\bibinfo {title} {Spin excitations and
  correlations in scanning tunneling spectroscopy},\ }\href
  {https://doi.org/10.1088/1367-2630/17/6/063016} {\bibfield  {journal}
  {\bibinfo  {journal} {New J. Phys.}\ }\textbf {\bibinfo {volume} {17}},\
  \bibinfo {pages} {063016} (\bibinfo {year} {2015})}\BibitemShut {NoStop}%
\bibitem [{\citenamefont {Heinrich}\ \emph {et~al.}(2004)\citenamefont
  {Heinrich}, \citenamefont {Gupta}, \citenamefont {Lutz},\ and\ \citenamefont
  {Eigler}}]{heinrich2004}%
  \BibitemOpen
  \bibfield  {author} {\bibinfo {author} {\bibfnamefont {A.~J.}\ \bibnamefont
  {Heinrich}}, \bibinfo {author} {\bibfnamefont {J.~A.}\ \bibnamefont {Gupta}},
  \bibinfo {author} {\bibfnamefont {C.~P.}\ \bibnamefont {Lutz}},\ and\
  \bibinfo {author} {\bibfnamefont {D.~M.}\ \bibnamefont {Eigler}},\ }\bibfield
   {title} {\bibinfo {title} {Single-atom spin-flip spectroscopy},\ }\href
  {https://doi.org/10.1126/science.1101077} {\bibfield  {journal} {\bibinfo
  {journal} {Science}\ }\textbf {\bibinfo {volume} {306}},\ \bibinfo {pages}
  {466} (\bibinfo {year} {2004})}\BibitemShut {NoStop}%
\bibitem [{\citenamefont {Hirjibehedin}\ \emph {et~al.}(2006)\citenamefont
  {Hirjibehedin}, \citenamefont {Lutz},\ and\ \citenamefont
  {Heinrich}}]{hirjibehedin2006}%
  \BibitemOpen
  \bibfield  {author} {\bibinfo {author} {\bibfnamefont {C.~F.}\ \bibnamefont
  {Hirjibehedin}}, \bibinfo {author} {\bibfnamefont {C.~P.}\ \bibnamefont
  {Lutz}},\ and\ \bibinfo {author} {\bibfnamefont {A.~J.}\ \bibnamefont
  {Heinrich}},\ }\bibfield  {title} {\bibinfo {title} {Spin coupling in
  engineered atomic structures},\ }\href
  {https://doi.org/10.1126/science.1125398} {\bibfield  {journal} {\bibinfo
  {journal} {Science}\ }\textbf {\bibinfo {volume} {312}},\ \bibinfo {pages}
  {1021} (\bibinfo {year} {2006})}\BibitemShut {NoStop}%
\bibitem [{\citenamefont {Pan}\ \emph {et~al.}(1998)\citenamefont {Pan},
  \citenamefont {Hudson},\ and\ \citenamefont {Davis}}]{Pan1998}%
  \BibitemOpen
  \bibfield  {author} {\bibinfo {author} {\bibfnamefont {S.~H.}\ \bibnamefont
  {Pan}}, \bibinfo {author} {\bibfnamefont {E.~W.}\ \bibnamefont {Hudson}},\
  and\ \bibinfo {author} {\bibfnamefont {J.~C.}\ \bibnamefont {Davis}},\
  }\bibfield  {title} {\bibinfo {title} {Vacuum tunneling of superconducting
  quasiparticles from atomically sharp scanning tunneling microscope tips},\
  }\href {https://doi.org/10.1063/1.122654} {\bibfield  {journal} {\bibinfo
  {journal} {Appl. Phys. Lett.}\ }\textbf {\bibinfo {volume} {73}},\ \bibinfo
  {pages} {2992} (\bibinfo {year} {1998})}\BibitemShut {NoStop}%
\bibitem [{\citenamefont {Ugeda}\ \emph {et~al.}(2015)\citenamefont {Ugeda},
  \citenamefont {Bradley}, \citenamefont {Zhang}, \citenamefont {Onishi},
  \citenamefont {Chen}, \citenamefont {Ruan}, \citenamefont
  {Ojeda-Aristizabal}, \citenamefont {Ryu}, \citenamefont {Edmonds},
  \citenamefont {Tsai}, \citenamefont {Riss}, \citenamefont {Mo}, \citenamefont
  {Lee}, \citenamefont {Zettl}, \citenamefont {Hussain}, \citenamefont {Shen},\
  and\ \citenamefont {Crommie}}]{ugeda2016}%
  \BibitemOpen
  \bibfield  {author} {\bibinfo {author} {\bibfnamefont {M.~M.}\ \bibnamefont
  {Ugeda}}, \bibinfo {author} {\bibfnamefont {A.~J.}\ \bibnamefont {Bradley}},
  \bibinfo {author} {\bibfnamefont {Y.}~\bibnamefont {Zhang}}, \bibinfo
  {author} {\bibfnamefont {S.}~\bibnamefont {Onishi}}, \bibinfo {author}
  {\bibfnamefont {Y.}~\bibnamefont {Chen}}, \bibinfo {author} {\bibfnamefont
  {W.}~\bibnamefont {Ruan}}, \bibinfo {author} {\bibfnamefont {C.}~\bibnamefont
  {Ojeda-Aristizabal}}, \bibinfo {author} {\bibfnamefont {H.}~\bibnamefont
  {Ryu}}, \bibinfo {author} {\bibfnamefont {M.~T.}\ \bibnamefont {Edmonds}},
  \bibinfo {author} {\bibfnamefont {H.-Z.}\ \bibnamefont {Tsai}}, \bibinfo
  {author} {\bibfnamefont {A.}~\bibnamefont {Riss}}, \bibinfo {author}
  {\bibfnamefont {S.-K.}\ \bibnamefont {Mo}}, \bibinfo {author} {\bibfnamefont
  {D.}~\bibnamefont {Lee}}, \bibinfo {author} {\bibfnamefont {A.}~\bibnamefont
  {Zettl}}, \bibinfo {author} {\bibfnamefont {Z.}~\bibnamefont {Hussain}},
  \bibinfo {author} {\bibfnamefont {Z.-X.}\ \bibnamefont {Shen}},\ and\
  \bibinfo {author} {\bibfnamefont {M.~F.}\ \bibnamefont {Crommie}},\
  }\bibfield  {title} {\bibinfo {title} {Characterization of collective ground
  states in single-layer {NbSe}$_2$},\ }\href
  {https://doi.org/10.1038/nphys3527} {\bibfield  {journal} {\bibinfo
  {journal} {Nat. Phys.}\ }\textbf {\bibinfo {volume} {12}},\ \bibinfo {pages}
  {92} (\bibinfo {year} {2015})}\BibitemShut {NoStop}%
\bibitem [{\citenamefont {Liebhaber}\ \emph {et~al.}(2019)\citenamefont
  {Liebhaber}, \citenamefont {Gonz{\'{a}}lez}, \citenamefont {Baba},
  \citenamefont {Reecht}, \citenamefont {Heinrich}, \citenamefont {Rohlf},
  \citenamefont {Rossnagel}, \citenamefont {von Oppen},\ and\ \citenamefont
  {Franke}}]{Liebhaber2019}%
  \BibitemOpen
  \bibfield  {author} {\bibinfo {author} {\bibfnamefont {E.}~\bibnamefont
  {Liebhaber}}, \bibinfo {author} {\bibfnamefont {S.~A.}\ \bibnamefont
  {Gonz{\'{a}}lez}}, \bibinfo {author} {\bibfnamefont {R.}~\bibnamefont
  {Baba}}, \bibinfo {author} {\bibfnamefont {G.}~\bibnamefont {Reecht}},
  \bibinfo {author} {\bibfnamefont {B.~W.}\ \bibnamefont {Heinrich}}, \bibinfo
  {author} {\bibfnamefont {S.}~\bibnamefont {Rohlf}}, \bibinfo {author}
  {\bibfnamefont {K.}~\bibnamefont {Rossnagel}}, \bibinfo {author}
  {\bibfnamefont {F.}~\bibnamefont {von Oppen}},\ and\ \bibinfo {author}
  {\bibfnamefont {K.~J.}\ \bibnamefont {Franke}},\ }\bibfield  {title}
  {\bibinfo {title} {{Yu-Shiba-Rusinov} states in the charge-density modulated
  superconductor {NbSe}$_2$},\ }\href
  {https://doi.org/10.1021/acs.nanolett.9b03988} {\bibfield  {journal}
  {\bibinfo  {journal} {Nano Lett.}\ }\textbf {\bibinfo {volume} {20}},\
  \bibinfo {pages} {339} (\bibinfo {year} {2019})}\BibitemShut {NoStop}%
\bibitem [{\citenamefont {Bork}\ \emph {et~al.}(2011)\citenamefont {Bork},
  \citenamefont {hui Zhang}, \citenamefont {Diekh\"{o}ner}, \citenamefont
  {Borda}, \citenamefont {Simon}, \citenamefont {Kroha}, \citenamefont {Wahl},\
  and\ \citenamefont {Kern}}]{bork2011}%
  \BibitemOpen
  \bibfield  {author} {\bibinfo {author} {\bibfnamefont {J.}~\bibnamefont
  {Bork}}, \bibinfo {author} {\bibfnamefont {Y.}~\bibnamefont {hui Zhang}},
  \bibinfo {author} {\bibfnamefont {L.}~\bibnamefont {Diekh\"{o}ner}}, \bibinfo
  {author} {\bibfnamefont {L.}~\bibnamefont {Borda}}, \bibinfo {author}
  {\bibfnamefont {P.}~\bibnamefont {Simon}}, \bibinfo {author} {\bibfnamefont
  {J.}~\bibnamefont {Kroha}}, \bibinfo {author} {\bibfnamefont
  {P.}~\bibnamefont {Wahl}},\ and\ \bibinfo {author} {\bibfnamefont
  {K.}~\bibnamefont {Kern}},\ }\bibfield  {title} {\bibinfo {title} {A tunable
  two-impurity {Kondo} system in an atomic point contact},\ }\href
  {https://doi.org/10.1038/nphys2076} {\bibfield  {journal} {\bibinfo
  {journal} {Nat. Phys.}\ }\textbf {\bibinfo {volume} {7}},\ \bibinfo {pages}
  {901} (\bibinfo {year} {2011})}\BibitemShut {NoStop}%
\bibitem [{\citenamefont {Yang}\ \emph {et~al.}(2019)\citenamefont {Yang},
  \citenamefont {Paul}, \citenamefont {Natterer}, \citenamefont {Lado},
  \citenamefont {Bae}, \citenamefont {Willke}, \citenamefont {Choi},
  \citenamefont {Ferr\'on}, \citenamefont {Fern\'andez-Rossier}, \citenamefont
  {Heinrich},\ and\ \citenamefont {Lutz}}]{kai2019}%
  \BibitemOpen
  \bibfield  {author} {\bibinfo {author} {\bibfnamefont {K.}~\bibnamefont
  {Yang}}, \bibinfo {author} {\bibfnamefont {W.}~\bibnamefont {Paul}}, \bibinfo
  {author} {\bibfnamefont {F.~D.}\ \bibnamefont {Natterer}}, \bibinfo {author}
  {\bibfnamefont {J.~L.}\ \bibnamefont {Lado}}, \bibinfo {author}
  {\bibfnamefont {Y.}~\bibnamefont {Bae}}, \bibinfo {author} {\bibfnamefont
  {P.}~\bibnamefont {Willke}}, \bibinfo {author} {\bibfnamefont
  {T.}~\bibnamefont {Choi}}, \bibinfo {author} {\bibfnamefont {A.}~\bibnamefont
  {Ferr\'on}}, \bibinfo {author} {\bibfnamefont {J.}~\bibnamefont
  {Fern\'andez-Rossier}}, \bibinfo {author} {\bibfnamefont {A.~J.}\
  \bibnamefont {Heinrich}},\ and\ \bibinfo {author} {\bibfnamefont {C.~P.}\
  \bibnamefont {Lutz}},\ }\bibfield  {title} {\bibinfo {title} {Tuning the
  exchange bias on a single atom from 1 {mT} to 10 {T}},\ }\href
  {https://doi.org/10.1103/PhysRevLett.122.227203} {\bibfield  {journal}
  {\bibinfo  {journal} {Phys. Rev. Lett.}\ }\textbf {\bibinfo {volume} {122}},\
  \bibinfo {pages} {227203} (\bibinfo {year} {2019})}\BibitemShut {NoStop}%
\bibitem [{SI()}]{SI}%
  \BibitemOpen
  \href@noop {} {\bibinfo {title} {See supplemental material, including
  references [38-44], for further details.}}\BibitemShut {Stop}%
\bibitem [{\citenamefont {White}(1992)}]{PhysRevLett.69.2863}%
  \BibitemOpen
  \bibfield  {author} {\bibinfo {author} {\bibfnamefont {S.~R.}\ \bibnamefont
  {White}},\ }\bibfield  {title} {\bibinfo {title} {Density matrix formulation
  for quantum renormalization groups},\ }\href
  {https://doi.org/10.1103/PhysRevLett.69.2863} {\bibfield  {journal} {\bibinfo
   {journal} {Phys. Rev. Lett.}\ }\textbf {\bibinfo {volume} {69}},\ \bibinfo
  {pages} {2863} (\bibinfo {year} {1992})}\BibitemShut {NoStop}%
\bibitem [{\citenamefont {Lado}\ and\ \citenamefont
  {Zilberberg}(2019)}]{PhysRevResearch.1.033009}%
  \BibitemOpen
  \bibfield  {author} {\bibinfo {author} {\bibfnamefont {J.~L.}\ \bibnamefont
  {Lado}}\ and\ \bibinfo {author} {\bibfnamefont {O.}~\bibnamefont
  {Zilberberg}},\ }\bibfield  {title} {\bibinfo {title} {Topological spin
  excitations in {Harper-Heisenberg} spin chains},\ }\href
  {https://doi.org/10.1103/PhysRevResearch.1.033009} {\bibfield  {journal}
  {\bibinfo  {journal} {Phys. Rev. Res.}\ }\textbf {\bibinfo {volume} {1}},\
  \bibinfo {pages} {033009} (\bibinfo {year} {2019})}\BibitemShut {NoStop}%
\bibitem [{\citenamefont {Lado}\ and\ \citenamefont
  {Sigrist}(2020)}]{PhysRevResearch.2.023347}%
  \BibitemOpen
  \bibfield  {author} {\bibinfo {author} {\bibfnamefont {J.~L.}\ \bibnamefont
  {Lado}}\ and\ \bibinfo {author} {\bibfnamefont {M.}~\bibnamefont {Sigrist}},\
  }\bibfield  {title} {\bibinfo {title} {Solitonic in-gap modes in a
  superconductor-quantum antiferromagnet interface},\ }\href
  {https://doi.org/10.1103/PhysRevResearch.2.023347} {\bibfield  {journal}
  {\bibinfo  {journal} {Phys. Rev. Res.}\ }\textbf {\bibinfo {volume} {2}},\
  \bibinfo {pages} {023347} (\bibinfo {year} {2020})}\BibitemShut {NoStop}%
\bibitem [{\citenamefont {Ganahl}\ \emph {et~al.}(2014)\citenamefont {Ganahl},
  \citenamefont {Thunstr\"om}, \citenamefont {Verstraete}, \citenamefont
  {Held},\ and\ \citenamefont {Evertz}}]{PhysRevB.90.045144}%
  \BibitemOpen
  \bibfield  {author} {\bibinfo {author} {\bibfnamefont {M.}~\bibnamefont
  {Ganahl}}, \bibinfo {author} {\bibfnamefont {P.}~\bibnamefont {Thunstr\"om}},
  \bibinfo {author} {\bibfnamefont {F.}~\bibnamefont {Verstraete}}, \bibinfo
  {author} {\bibfnamefont {K.}~\bibnamefont {Held}},\ and\ \bibinfo {author}
  {\bibfnamefont {H.~G.}\ \bibnamefont {Evertz}},\ }\bibfield  {title}
  {\bibinfo {title} {Chebyshev expansion for impurity models using matrix
  product states},\ }\href {https://doi.org/10.1103/PhysRevB.90.045144}
  {\bibfield  {journal} {\bibinfo  {journal} {Phys. Rev. B}\ }\textbf {\bibinfo
  {volume} {90}},\ \bibinfo {pages} {045144} (\bibinfo {year}
  {2014})}\BibitemShut {NoStop}%
\bibitem [{\citenamefont {Wolf}\ \emph {et~al.}(2014)\citenamefont {Wolf},
  \citenamefont {McCulloch}, \citenamefont {Parcollet},\ and\ \citenamefont
  {Schollw\"ock}}]{PhysRevB.90.115124}%
  \BibitemOpen
  \bibfield  {author} {\bibinfo {author} {\bibfnamefont {F.~A.}\ \bibnamefont
  {Wolf}}, \bibinfo {author} {\bibfnamefont {I.~P.}\ \bibnamefont {McCulloch}},
  \bibinfo {author} {\bibfnamefont {O.}~\bibnamefont {Parcollet}},\ and\
  \bibinfo {author} {\bibfnamefont {U.}~\bibnamefont {Schollw\"ock}},\
  }\bibfield  {title} {\bibinfo {title} {Chebyshev matrix product state
  impurity solver for dynamical mean-field theory},\ }\href
  {https://doi.org/10.1103/PhysRevB.90.115124} {\bibfield  {journal} {\bibinfo
  {journal} {Phys. Rev. B}\ }\textbf {\bibinfo {volume} {90}},\ \bibinfo
  {pages} {115124} (\bibinfo {year} {2014})}\BibitemShut {NoStop}%
\bibitem [{\citenamefont {Wei\ss{}e}\ \emph {et~al.}(2006)\citenamefont
  {Wei\ss{}e}, \citenamefont {Wellein}, \citenamefont {Alvermann},\ and\
  \citenamefont {Fehske}}]{RevModPhys.78.275}%
  \BibitemOpen
  \bibfield  {author} {\bibinfo {author} {\bibfnamefont {A.}~\bibnamefont
  {Wei\ss{}e}}, \bibinfo {author} {\bibfnamefont {G.}~\bibnamefont {Wellein}},
  \bibinfo {author} {\bibfnamefont {A.}~\bibnamefont {Alvermann}},\ and\
  \bibinfo {author} {\bibfnamefont {H.}~\bibnamefont {Fehske}},\ }\bibfield
  {title} {\bibinfo {title} {The kernel polynomial method},\ }\href
  {https://doi.org/10.1103/RevModPhys.78.275} {\bibfield  {journal} {\bibinfo
  {journal} {Rev. Mod. Phys.}\ }\textbf {\bibinfo {volume} {78}},\ \bibinfo
  {pages} {275} (\bibinfo {year} {2006})}\BibitemShut {NoStop}%
\bibitem [{\citenamefont {{Karjalainen}}\ \emph {et~al.}(2022)\citenamefont
  {{Karjalainen}}, \citenamefont {{Lippo}}, \citenamefont {{Chen}},
  \citenamefont {{Koch}}, \citenamefont {{Fumega}},\ and\ \citenamefont
  {{Lado}}}]{2022arXiv221207893K}%
  \BibitemOpen
  \bibfield  {author} {\bibinfo {author} {\bibfnamefont {N.}~\bibnamefont
  {{Karjalainen}}}, \bibinfo {author} {\bibfnamefont {Z.}~\bibnamefont
  {{Lippo}}}, \bibinfo {author} {\bibfnamefont {G.}~\bibnamefont {{Chen}}},
  \bibinfo {author} {\bibfnamefont {R.}~\bibnamefont {{Koch}}}, \bibinfo
  {author} {\bibfnamefont {A.~O.}\ \bibnamefont {{Fumega}}},\ and\ \bibinfo
  {author} {\bibfnamefont {J.~L.}\ \bibnamefont {{Lado}}},\ }\bibfield  {title}
  {\bibinfo {title} {{Hamiltonian inference from dynamical excitations in spin
  quantum dots}},\ }\href@noop {} {\bibfield  {journal} {\bibinfo  {journal}
  {arXiv e-prints}\ ,\ \bibinfo {eid} {arXiv:2212.07893}} (\bibinfo {year}
  {2022})},\ \Eprint {https://arxiv.org/abs/2212.07893} {arXiv:2212.07893
  [cond-mat.mes-hall]} \BibitemShut {NoStop}%
\bibitem [{\citenamefont {Wang}\ \emph {et~al.}(1993)\citenamefont {Wang},
  \citenamefont {Wu},\ and\ \citenamefont {Freeman}}]{wang1993}%
  \BibitemOpen
  \bibfield  {author} {\bibinfo {author} {\bibfnamefont {D.-s.}\ \bibnamefont
  {Wang}}, \bibinfo {author} {\bibfnamefont {R.}~\bibnamefont {Wu}},\ and\
  \bibinfo {author} {\bibfnamefont {A.~J.}\ \bibnamefont {Freeman}},\
  }\bibfield  {title} {\bibinfo {title} {First-principles theory of surface
  magnetocrystalline anisotropy and the diatomic-pair model},\ }\href
  {https://doi.org/10.1103/PhysRevB.47.14932} {\bibfield  {journal} {\bibinfo
  {journal} {Phys. Rev. B}\ }\textbf {\bibinfo {volume} {47}},\ \bibinfo
  {pages} {14932} (\bibinfo {year} {1993})}\BibitemShut {NoStop}%
\bibitem [{\citenamefont {Gambardella}\ \emph {et~al.}(2009)\citenamefont
  {Gambardella}, \citenamefont {Stepanow}, \citenamefont {Dmitriev},
  \citenamefont {Honolka}, \citenamefont {de~Groot}, \citenamefont
  {Lingenfelder}, \citenamefont {Gupta}, \citenamefont {Sarma}, \citenamefont
  {Bencok}, \citenamefont {Stanescu}, \citenamefont {Clair}, \citenamefont
  {Pons}, \citenamefont {Lin}, \citenamefont {Seitsonen}, \citenamefont
  {Brune}, \citenamefont {Barth},\ and\ \citenamefont
  {Kern}}]{gambardella2009}%
  \BibitemOpen
  \bibfield  {author} {\bibinfo {author} {\bibfnamefont {P.}~\bibnamefont
  {Gambardella}}, \bibinfo {author} {\bibfnamefont {S.}~\bibnamefont
  {Stepanow}}, \bibinfo {author} {\bibfnamefont {A.}~\bibnamefont {Dmitriev}},
  \bibinfo {author} {\bibfnamefont {J.}~\bibnamefont {Honolka}}, \bibinfo
  {author} {\bibfnamefont {F.~M.~F.}\ \bibnamefont {de~Groot}}, \bibinfo
  {author} {\bibfnamefont {M.}~\bibnamefont {Lingenfelder}}, \bibinfo {author}
  {\bibfnamefont {S.~S.}\ \bibnamefont {Gupta}}, \bibinfo {author}
  {\bibfnamefont {D.~D.}\ \bibnamefont {Sarma}}, \bibinfo {author}
  {\bibfnamefont {P.}~\bibnamefont {Bencok}}, \bibinfo {author} {\bibfnamefont
  {S.}~\bibnamefont {Stanescu}}, \bibinfo {author} {\bibfnamefont
  {S.}~\bibnamefont {Clair}}, \bibinfo {author} {\bibfnamefont
  {S.}~\bibnamefont {Pons}}, \bibinfo {author} {\bibfnamefont {N.}~\bibnamefont
  {Lin}}, \bibinfo {author} {\bibfnamefont {A.~P.}\ \bibnamefont {Seitsonen}},
  \bibinfo {author} {\bibfnamefont {H.}~\bibnamefont {Brune}}, \bibinfo
  {author} {\bibfnamefont {J.~V.}\ \bibnamefont {Barth}},\ and\ \bibinfo
  {author} {\bibfnamefont {K.}~\bibnamefont {Kern}},\ }\bibfield  {title}
  {\bibinfo {title} {Supramolecular control of the magnetic anisotropy
  in~two-dimensional high-spin {Fe} arrays at a metal~interface},\ }\href
  {https://doi.org/10.1038/nmat2376} {\bibfield  {journal} {\bibinfo  {journal}
  {Nat. Mater.}\ }\textbf {\bibinfo {volume} {8}},\ \bibinfo {pages} {189}
  (\bibinfo {year} {2009})}\BibitemShut {NoStop}%
\bibitem [{\citenamefont {Heinrich}\ \emph {et~al.}(2015)\citenamefont
  {Heinrich}, \citenamefont {Braun}, \citenamefont {Pascual},\ and\
  \citenamefont {Franke}}]{heinrich2015}%
  \BibitemOpen
  \bibfield  {author} {\bibinfo {author} {\bibfnamefont {B.~W.}\ \bibnamefont
  {Heinrich}}, \bibinfo {author} {\bibfnamefont {L.}~\bibnamefont {Braun}},
  \bibinfo {author} {\bibfnamefont {J.~I.}\ \bibnamefont {Pascual}},\ and\
  \bibinfo {author} {\bibfnamefont {K.~J.}\ \bibnamefont {Franke}},\ }\bibfield
   {title} {\bibinfo {title} {Tuning the magnetic anisotropy of single
  molecules},\ }\href {https://doi.org/10.1021/acs.nanolett.5b00987} {\bibfield
   {journal} {\bibinfo  {journal} {Nano Lett.}\ }\textbf {\bibinfo {volume}
  {15}},\ \bibinfo {pages} {4024} (\bibinfo {year} {2015})}\BibitemShut
  {NoStop}%
\bibitem [{dmr()}]{dmrgpy}%
  \BibitemOpen
  \href@noop {} {\bibinfo  {journal} {\mbox{DMRGpy Library}
  https://github.com/joselado/dmrgpy}\ }\BibitemShut {NoStop}%
\bibitem [{ITe()}]{ITensor}%
  \BibitemOpen
\bibfield  {journal} {  }\href@noop {} {\bibinfo  {journal} {\mbox{ITensor
  Library} http://itensor.org}\ }\BibitemShut {NoStop}%
\bibitem [{\citenamefont {Fishman}\ \emph {et~al.}(2022)\citenamefont
  {Fishman}, \citenamefont {White},\ and\ \citenamefont
  {Stoudenmire}}]{10.21468/SciPostPhysCodeb.4}%
  \BibitemOpen
\bibfield  {journal} {  }\bibfield  {author} {\bibinfo {author} {\bibfnamefont
  {M.}~\bibnamefont {Fishman}}, \bibinfo {author} {\bibfnamefont {S.~R.}\
  \bibnamefont {White}},\ and\ \bibinfo {author} {\bibfnamefont {E.~M.}\
  \bibnamefont {Stoudenmire}},\ }\bibfield  {title} {\bibinfo {title} {The
  {ITensor} software library for tensor network calculations},\ }\href
  {https://doi.org/10.21468/SciPostPhysCodeb.4} {\bibfield  {journal} {\bibinfo
   {journal} {SciPost Phys. Codebases}\ ,\ \bibinfo {pages} {4}} (\bibinfo
  {year} {2022})}\BibitemShut {NoStop}%
\bibitem [{\citenamefont {Ram}\ and\ \citenamefont
  {Kumar}(2017)}]{PhysRevB.96.075115}%
  \BibitemOpen
  \bibfield  {author} {\bibinfo {author} {\bibfnamefont {P.}~\bibnamefont
  {Ram}}\ and\ \bibinfo {author} {\bibfnamefont {B.}~\bibnamefont {Kumar}},\
  }\bibfield  {title} {\bibinfo {title} {Theory of quantum oscillations of
  magnetization in kondo insulators},\ }\href
  {https://doi.org/10.1103/PhysRevB.96.075115} {\bibfield  {journal} {\bibinfo
  {journal} {Phys. Rev. B}\ }\textbf {\bibinfo {volume} {96}},\ \bibinfo
  {pages} {075115} (\bibinfo {year} {2017})}\BibitemShut {NoStop}%
\bibitem [{\citenamefont {Ram}\ and\ \citenamefont
  {Kumar}(2019)}]{PhysRevB.99.235130}%
  \BibitemOpen
  \bibfield  {author} {\bibinfo {author} {\bibfnamefont {P.}~\bibnamefont
  {Ram}}\ and\ \bibinfo {author} {\bibfnamefont {B.}~\bibnamefont {Kumar}},\
  }\bibfield  {title} {\bibinfo {title} {Inversion and magnetic quantum
  oscillations in the symmetric periodic {Anderson} model},\ }\href
  {https://doi.org/10.1103/PhysRevB.99.235130} {\bibfield  {journal} {\bibinfo
  {journal} {Phys. Rev. B}\ }\textbf {\bibinfo {volume} {99}},\ \bibinfo
  {pages} {235130} (\bibinfo {year} {2019})}\BibitemShut {NoStop}%
\bibitem [{\citenamefont {{Chen}}\ \emph {et~al.}(2023)\citenamefont {{Chen}},
  \citenamefont {{Stoudenmire}}, \citenamefont {{Komijani}},\ and\
  \citenamefont {{Coleman}}}]{2023arXiv230209701C}%
  \BibitemOpen
  \bibfield  {author} {\bibinfo {author} {\bibfnamefont {J.}~\bibnamefont
  {{Chen}}}, \bibinfo {author} {\bibfnamefont {E.~M.}\ \bibnamefont
  {{Stoudenmire}}}, \bibinfo {author} {\bibfnamefont {Y.}~\bibnamefont
  {{Komijani}}},\ and\ \bibinfo {author} {\bibfnamefont {P.}~\bibnamefont
  {{Coleman}}},\ }\bibfield  {title} {\bibinfo {title} {Matrix product study of
  spin fractionalization in the {1D Kondo} insulator},\ }\href
  {https://doi.org/10.48550/arXiv.2302.09701} {\bibfield  {journal} {\bibinfo
  {journal} {arXiv e-prints}\ ,\ \bibinfo {eid} {arXiv:2302.09701}} (\bibinfo
  {year} {2023})},\ \Eprint {https://arxiv.org/abs/2302.09701}
  {arXiv:2302.09701 [cond-mat.str-el]} \BibitemShut {NoStop}%
\end{thebibliography}

\begin{thebibliography}{12}%
\makeatletter
\providecommand \@ifxundefined [1]{%
 \@ifx{#1\undefined}
}%
\providecommand \@ifnum [1]{%
 \ifnum #1\expandafter \@firstoftwo
 \else \expandafter \@secondoftwo
 \fi
}%
\providecommand \@ifx [1]{%
 \ifx #1\expandafter \@firstoftwo
 \else \expandafter \@secondoftwo
 \fi
}%
\providecommand \natexlab [1]{#1}%
\providecommand \enquote  [1]{``#1''}%
\providecommand \bibnamefont  [1]{#1}%
\providecommand \bibfnamefont [1]{#1}%
\providecommand \citenamefont [1]{#1}%
\providecommand \href@noop [0]{\@secondoftwo}%
\providecommand \href [0]{\begingroup \@sanitize@url \@href}%
\providecommand \@href[1]{\@@startlink{#1}\@@href}%
\providecommand \@@href[1]{\endgroup#1\@@endlink}%
\providecommand \@sanitize@url [0]{\catcode `\\12\catcode `\$12\catcode
  `\&12\catcode `\#12\catcode `\^12\catcode `\_12\catcode `\%12\relax}%
\providecommand \@@startlink[1]{}%
\providecommand \@@endlink[0]{}%
\providecommand \url  [0]{\begingroup\@sanitize@url \@url }%
\providecommand \@url [1]{\endgroup\@href {#1}{\urlprefix }}%
\providecommand \urlprefix  [0]{URL }%
\providecommand \Eprint [0]{\href }%
\providecommand \doibase [0]{https://doi.org/}%
\providecommand \selectlanguage [0]{\@gobble}%
\providecommand \bibinfo  [0]{\@secondoftwo}%
\providecommand \bibfield  [0]{\@secondoftwo}%
\providecommand \translation [1]{[#1]}%
\providecommand \BibitemOpen [0]{}%
\providecommand \bibitemStop [0]{}%
\providecommand \bibitemNoStop [0]{.\EOS\space}%
\providecommand \EOS [0]{\spacefactor3000\relax}%
\providecommand \BibitemShut  [1]{\csname bibitem#1\endcsname}%
\let\auto@bib@innerbib\@empty
\bibitem [{\citenamefont {Kezilebieke}\ \emph {et~al.}(2018)\citenamefont
  {Kezilebieke}, \citenamefont {Dvorak}, \citenamefont {Ojanen},\ and\
  \citenamefont {Liljeroth}}]{Kezilebieke2018}%
  \BibitemOpen
  \bibfield  {author} {\bibinfo {author} {\bibfnamefont {S.}~\bibnamefont
  {Kezilebieke}}, \bibinfo {author} {\bibfnamefont {M.}~\bibnamefont {Dvorak}},
  \bibinfo {author} {\bibfnamefont {T.}~\bibnamefont {Ojanen}},\ and\ \bibinfo
  {author} {\bibfnamefont {P.}~\bibnamefont {Liljeroth}},\ }\bibfield  {title}
  {\bibinfo {title} {Coupled {Yu-Shiba-Rusinov} states in molecular dimers on
  {NbSe}$_2$},\ }\href {https://doi.org/10.1021/acs.nanolett.7b05050}
  {\bibfield  {journal} {\bibinfo  {journal} {Nano Lett.}\ }\textbf {\bibinfo
  {volume} {18}},\ \bibinfo {pages} {2311} (\bibinfo {year}
  {2018})}\BibitemShut {NoStop}%
\bibitem [{\citenamefont {Wang}\ \emph {et~al.}(2021)\citenamefont {Wang},
  \citenamefont {Arabi}, \citenamefont {Kern},\ and\ \citenamefont
  {Ternes}}]{wang2021}%
  \BibitemOpen
  \bibfield  {author} {\bibinfo {author} {\bibfnamefont {Y.}~\bibnamefont
  {Wang}}, \bibinfo {author} {\bibfnamefont {S.}~\bibnamefont {Arabi}},
  \bibinfo {author} {\bibfnamefont {K.}~\bibnamefont {Kern}},\ and\ \bibinfo
  {author} {\bibfnamefont {M.}~\bibnamefont {Ternes}},\ }\bibfield  {title}
  {\bibinfo {title} {Symmetry mediated tunable molecular magnetism on a {2D}
  material},\ }\href {https://doi.org/10.1038/s42005-021-00601-8} {\bibfield
  {journal} {\bibinfo  {journal} {Commun. Phys.}\ }\textbf {\bibinfo {volume}
  {4}},\ \bibinfo {pages} {103} (\bibinfo {year} {2021})}\BibitemShut {NoStop}%
\bibitem [{dmr()}]{dmrgpy}%
  \BibitemOpen
  \href@noop {} {\bibinfo  {journal} {\mbox{DMRGpy Library}
  https://github.com/joselado/dmrgpy}\ }\BibitemShut {NoStop}%
\bibitem [{ITe()}]{ITensor}%
  \BibitemOpen
\bibfield  {journal} {  }\href@noop {} {\bibinfo  {journal} {\mbox{ITensor
  Library} http://itensor.org}\ }\BibitemShut {NoStop}%
\bibitem [{\citenamefont {Fishman}\ \emph {et~al.}(2022)\citenamefont
  {Fishman}, \citenamefont {White},\ and\ \citenamefont
  {Stoudenmire}}]{10.21468/SciPostPhysCodeb.4}%
  \BibitemOpen
\bibfield  {journal} {  }\bibfield  {author} {\bibinfo {author} {\bibfnamefont
  {M.}~\bibnamefont {Fishman}}, \bibinfo {author} {\bibfnamefont {S.~R.}\
  \bibnamefont {White}},\ and\ \bibinfo {author} {\bibfnamefont {E.~M.}\
  \bibnamefont {Stoudenmire}},\ }\bibfield  {title} {\bibinfo {title} {{The
  ITensor Software Library for Tensor Network Calculations}},\ }\href
  {https://doi.org/10.21468/SciPostPhysCodeb.4} {\bibfield  {journal} {\bibinfo
   {journal} {SciPost Phys. Codebases}\ ,\ \bibinfo {pages} {4}} (\bibinfo
  {year} {2022})}\BibitemShut {NoStop}%
\bibitem [{\citenamefont {White}(1992)}]{PhysRevLett.69.2863}%
  \BibitemOpen
  \bibfield  {author} {\bibinfo {author} {\bibfnamefont {S.~R.}\ \bibnamefont
  {White}},\ }\bibfield  {title} {\bibinfo {title} {Density matrix formulation
  for quantum renormalization groups},\ }\href
  {https://doi.org/10.1103/PhysRevLett.69.2863} {\bibfield  {journal} {\bibinfo
   {journal} {Phys. Rev. Lett.}\ }\textbf {\bibinfo {volume} {69}},\ \bibinfo
  {pages} {2863} (\bibinfo {year} {1992})}\BibitemShut {NoStop}%
\bibitem [{\citenamefont {Lado}\ and\ \citenamefont
  {Zilberberg}(2019)}]{PhysRevResearch.1.033009}%
  \BibitemOpen
  \bibfield  {author} {\bibinfo {author} {\bibfnamefont {J.~L.}\ \bibnamefont
  {Lado}}\ and\ \bibinfo {author} {\bibfnamefont {O.}~\bibnamefont
  {Zilberberg}},\ }\bibfield  {title} {\bibinfo {title} {Topological spin
  excitations in {Harper-Heisenberg} spin chains},\ }\href
  {https://doi.org/10.1103/PhysRevResearch.1.033009} {\bibfield  {journal}
  {\bibinfo  {journal} {Phys. Rev. Res.}\ }\textbf {\bibinfo {volume} {1}},\
  \bibinfo {pages} {033009} (\bibinfo {year} {2019})}\BibitemShut {NoStop}%
\bibitem [{\citenamefont {Lado}\ and\ \citenamefont
  {Sigrist}(2020)}]{PhysRevResearch.2.023347}%
  \BibitemOpen
  \bibfield  {author} {\bibinfo {author} {\bibfnamefont {J.~L.}\ \bibnamefont
  {Lado}}\ and\ \bibinfo {author} {\bibfnamefont {M.}~\bibnamefont {Sigrist}},\
  }\bibfield  {title} {\bibinfo {title} {Solitonic in-gap modes in a
  superconductor-quantum antiferromagnet interface},\ }\href
  {https://doi.org/10.1103/PhysRevResearch.2.023347} {\bibfield  {journal}
  {\bibinfo  {journal} {Phys. Rev. Res.}\ }\textbf {\bibinfo {volume} {2}},\
  \bibinfo {pages} {023347} (\bibinfo {year} {2020})}\BibitemShut {NoStop}%
\bibitem [{\citenamefont {Ganahl}\ \emph {et~al.}(2014)\citenamefont {Ganahl},
  \citenamefont {Thunstr\"om}, \citenamefont {Verstraete}, \citenamefont
  {Held},\ and\ \citenamefont {Evertz}}]{PhysRevB.90.045144}%
  \BibitemOpen
  \bibfield  {author} {\bibinfo {author} {\bibfnamefont {M.}~\bibnamefont
  {Ganahl}}, \bibinfo {author} {\bibfnamefont {P.}~\bibnamefont {Thunstr\"om}},
  \bibinfo {author} {\bibfnamefont {F.}~\bibnamefont {Verstraete}}, \bibinfo
  {author} {\bibfnamefont {K.}~\bibnamefont {Held}},\ and\ \bibinfo {author}
  {\bibfnamefont {H.~G.}\ \bibnamefont {Evertz}},\ }\bibfield  {title}
  {\bibinfo {title} {Chebyshev expansion for impurity models using matrix
  product states},\ }\href {https://doi.org/10.1103/PhysRevB.90.045144}
  {\bibfield  {journal} {\bibinfo  {journal} {Phys. Rev. B}\ }\textbf {\bibinfo
  {volume} {90}},\ \bibinfo {pages} {045144} (\bibinfo {year}
  {2014})}\BibitemShut {NoStop}%
\bibitem [{\citenamefont {Wolf}\ \emph {et~al.}(2014)\citenamefont {Wolf},
  \citenamefont {McCulloch}, \citenamefont {Parcollet},\ and\ \citenamefont
  {Schollw\"ock}}]{PhysRevB.90.115124}%
  \BibitemOpen
  \bibfield  {author} {\bibinfo {author} {\bibfnamefont {F.~A.}\ \bibnamefont
  {Wolf}}, \bibinfo {author} {\bibfnamefont {I.~P.}\ \bibnamefont {McCulloch}},
  \bibinfo {author} {\bibfnamefont {O.}~\bibnamefont {Parcollet}},\ and\
  \bibinfo {author} {\bibfnamefont {U.}~\bibnamefont {Schollw\"ock}},\
  }\bibfield  {title} {\bibinfo {title} {Chebyshev matrix product state
  impurity solver for dynamical mean-field theory},\ }\href
  {https://doi.org/10.1103/PhysRevB.90.115124} {\bibfield  {journal} {\bibinfo
  {journal} {Phys. Rev. B}\ }\textbf {\bibinfo {volume} {90}},\ \bibinfo
  {pages} {115124} (\bibinfo {year} {2014})}\BibitemShut {NoStop}%
\bibitem [{\citenamefont {Wei\ss{}e}\ \emph {et~al.}(2006)\citenamefont
  {Wei\ss{}e}, \citenamefont {Wellein}, \citenamefont {Alvermann},\ and\
  \citenamefont {Fehske}}]{RevModPhys.78.275}%
  \BibitemOpen
  \bibfield  {author} {\bibinfo {author} {\bibfnamefont {A.}~\bibnamefont
  {Wei\ss{}e}}, \bibinfo {author} {\bibfnamefont {G.}~\bibnamefont {Wellein}},
  \bibinfo {author} {\bibfnamefont {A.}~\bibnamefont {Alvermann}},\ and\
  \bibinfo {author} {\bibfnamefont {H.}~\bibnamefont {Fehske}},\ }\bibfield
  {title} {\bibinfo {title} {The kernel polynomial method},\ }\href
  {https://doi.org/10.1103/RevModPhys.78.275} {\bibfield  {journal} {\bibinfo
  {journal} {Rev. Mod. Phys.}\ }\textbf {\bibinfo {volume} {78}},\ \bibinfo
  {pages} {275} (\bibinfo {year} {2006})}\BibitemShut {NoStop}%
\bibitem [{\citenamefont {{Karjalainen}}\ \emph {et~al.}(2022)\citenamefont
  {{Karjalainen}}, \citenamefont {{Lippo}}, \citenamefont {{Chen}},
  \citenamefont {{Koch}}, \citenamefont {{Fumega}},\ and\ \citenamefont
  {{Lado}}}]{2022arXiv221207893K}%
  \BibitemOpen
  \bibfield  {author} {\bibinfo {author} {\bibfnamefont {N.}~\bibnamefont
  {{Karjalainen}}}, \bibinfo {author} {\bibfnamefont {Z.}~\bibnamefont
  {{Lippo}}}, \bibinfo {author} {\bibfnamefont {G.}~\bibnamefont {{Chen}}},
  \bibinfo {author} {\bibfnamefont {R.}~\bibnamefont {{Koch}}}, \bibinfo
  {author} {\bibfnamefont {A.~O.}\ \bibnamefont {{Fumega}}},\ and\ \bibinfo
  {author} {\bibfnamefont {J.~L.}\ \bibnamefont {{Lado}}},\ }\bibfield  {title}
  {\bibinfo {title} {{Hamiltonian inference from dynamical excitations in spin
  quantum dots}},\ }\href@noop {} {\bibfield  {journal} {\bibinfo  {journal}
  {arXiv e-prints}\ ,\ \bibinfo {eid} {arXiv:2212.07893}} (\bibinfo {year}
  {2022})},\ \Eprint {https://arxiv.org/abs/2212.07893} {arXiv:2212.07893
  [cond-mat.mes-hall]} \BibitemShut {NoStop}%
\end{thebibliography}
%

\end{document}